\documentclass[preprint,aps,pre]{revtex4}

\usepackage{graphicx}

\begin{document}

\title
{Dynamical stability of the weakly nonharmonic\\
 propeller-shaped planar  Brownian rotator}

\author{Igor \surname{Petrovi\' c}}\email{igorpetrovicsb@gmail.com}
\affiliation{Svetog Save 98, 18230 Sokobanja, Serbia}
\author{Jasmina \surname{Jekni\' c-Dugi\' c}}\email{jjeknic@pmf.ni.ac.rs}
\affiliation{University of Ni\v s, Faculty of Science and Mathematics, Vi\v
segradska 33, 18000 Ni\v s, Serbia}
\author{Momir \surname{Arsenijevi\'c}}\email{fajnman@gmail.com}
\affiliation{University of Kragujevac, Faculty of Science, Radoja
Domanovi\' ca 12, 34000 Kragujevac, Serbia}
\author{Miroljub \surname{Dugi\'c}}\email{mdugic18@sbb.rs}
\affiliation{University of Kragujevac, Faculty of Science, Radoja
Domanovi\' ca 12, 34000 Kragujevac, Serbia}

\date{\today}

\begin{abstract}
Dynamical stability is a prerequisite for  control and functioning of  desired nano-machines. We utilize the Caldeira-Leggett master equation to investigate dynamical stability of  molecular cogwheels
modeled as a rigid, propeller-shaped planar rotator. In order to match certain expected realistic physical situations, we consider a weakly  nonharmonic external potential for the rotator.
Two methods for investigating stability are used. First, we employ
a quantum-mechanical counterpart of the so-called "first passage time" method.
Second, we investigate time dependence of the standard deviation of the rotator for both the angle and  angular momentum quantum observables. A perturbation-like procedure is introduced and implemented in order to provide the closed set of  differential equations for the moments. Extensive analysis  is performed for different combinations of the values of system parameters. The two methods are, in a sense,  mutually complementary. Appropriate for the short time behavior, the first passage time  exhibits a numerically-relevant dependence only on the damping factor as well as on the rotator size. On the other hand, the standard deviations for both the angle and angular momentum observables exhibit strong dependence on the parameter values for both short and long time intervals. Contrary to our expectations,  the time decrease of the standard deviations is found for certain parameter regimes. In addition, for certain parameter regimes nonmonotonic dependence on the rotator size is observed for the standard deviations and for the damping of the oscillation amplitude. Hence non-fulfillment of the classical expectation that the size of the rotator can be reduced to the inertia of the rotator. In effect, the task of designing the desired protocols for the proper control of the molecular rotations becomes an optimization problem that requires further technical elaboration.
\end{abstract}
%\pacs{05.40.Fb, 64.60.De, 64.60.al, 64.60.ae}

\maketitle

\section{Introduction}
\label{uvod}

Functional parts of the realistic and  desired nano-machines have  a finite spatial size and  definite geometrical shape as well as
exposed to  environmental influence \cite{jones,kottas,goodsell,kudernec}. A desired function of those parts {\it determines} their size and geometry (geometrical shape), which, in turn,  determines the
environmental influence. This  poses a challenge for both  theoretical studies and experimental investigations of the {\it realistic} nano-scale systems.
Particularly, description of nano-sized rotating molecules monitored by the many-particle environment
"cannot be reduced to any of the previously known impurity problems of condensed matter physics" \cite{schmidt}.

There is not yet a general quantum theory to link {\it dynamics} with the system's {\it spatial size and geometry}. Certain simplified models  (such as e.g. the sphere- or ellipsoid- or a rod-like shaped rotators) \cite{chandrasekar,tenhagen} are
typically considered to have rotational symmetry in regard to the  system of interest as well as  the homogeneous environment. The requirement of rotational symmetry justifies, for some models, construction of the effective, rotationally symmetric interactions or external potentials for the rotator system \cite{roulet}. The related quantum master equations  are typically assumed or constructed to be Markovian \cite{roulet,stickler,papendell,stickler2}, that is, of the Lindblad form \cite{breuer,rivas}.

For larger molecular species with  high temperature of the environment it is expected that the classical theory may work well. Therefore, one may ask whether the quantum-mechanical description may be of any practical use. Nevertheless, there are observable effects  such as the so-called barriers to rotation that, in some cases, require the quantum-mechanical description \cite{kottas}. Certain quantum corrections  are found and deeply investigated for some analogous classical models  \cite{melnikov,coffey1,coffey2,coffey3,coffey4}.
In certain scenarios, the individually negligible quantum-mechanical contributions may accumulate to such extent that the classical theory is of  limited use \cite{JJD1}. To this end, the standard classical theory of the Brownian rotator does not support the rotational symmetry \cite{kottas}. E.g., the external electric field applied to  dipolar molecules as well as  the molecular rotators resting on the solid surface may introduce (in the zeroth approximation) the external harmonic field for the rotator of the form $I\omega^2\varphi^2/2$ for a molecule with the moment of inertia $I$, the circular frequency $\omega$ and  the azimuthal angle of rotation $\varphi$ \cite{kottas}. For such scenarios, the rotational $2\pi$-symmetry is not applicable. Then, the standard quantization procedure \cite{JJD1,japs} distinguishes the  Caldeira-Leggett master equation \cite{breuer,CL} as a possibly useful model for the realistic physical situations.
A semiclassical master equation for the Brownian particle's Wigner function appears as an alternative method that is
 equipped with the powerful calculation tools \cite{coffey1,coffey2,coffey3,coffey4}.

Analysis of the rotator's size and shape can be performed particularly for the propeller-like molecular rotators in the context of certain plausible (often used) assumptions regarding the rotator's interaction with the thermal bath \cite{JJD1}.   Then the rotator's size can be introduced by the linear dependence of both the moment of inertia and the strength of interaction (and therefore of the damping factor) on the number $N$ of the blades of the propeller rotator \cite{JJD1}. Those linearities do not appear for the general case, e.g., for the mutually dependent blades [in which case the local environments may also become mutually dependent]. Dynamical stability has been investigated for the free rotator and  rotator in the external harmonic potential [14]. The absence of  simple rules or recipes for utilizing the rotator stability is acknowledged \cite{JJD1}.

Proceeding now we utilize the Caldeira-Legget master equation to investigate stability of a quantum rotator, which is placed in the external potential of the form $V(\hat\varphi) = I\omega^2\hat\varphi^2/2 - b \hat\varphi^3$ with the small real parameter $b>0$. Our task is to investigate dynamical stability of the rotator of the size (the number of blades) $N$ with the average moment of inertia  $I_{\circ}$ and the damping factor $\gamma_{\circ}$. Quantum mechanical consistency of the  model that does not account for the problematic uncertainty relation for the angle and  angular momentum observables allows only small rotations to be considered \cite{breitenberger,deck}.
The physical origin of the cubic term may lie in the external driving field as well as in the effective intrinsic potential for the rotator.
Otherwise, the use of this kind of  potential can be found in investigations of certain nonlinear dynamical systems, notably the dynamics of the initial metastable state (classical or quantum) regarding, e.g., the so-called ''noise enhanced stability'' effect \cite{dubkov,spagnolo1,spagnolo2,spagnolo3}, non-linear friction models \cite{burada}, decay of  unstable states \cite{ornigotti}, stabilization of volatility in financial market \cite{valenti} and a model of certain chemical reactions \cite{coffey1}.

We do not restrict our considerations to the original \cite{CL} assumption of the weak coupling and the high temperature of the environment. Rather, we regard the Caldeira-Leggett master equation in the phenomenological  sense  as emphasized in \cite{ferialdi}.

The various approaches  are developed to describe  decay of unstable states.
Notably, the escape from a metastable state (the "Kramers problem") and dynamics of the standard deviations for the relevant variables are of particular interest.
Escape from a metastable state (from a potential well) can be described in the mutually nonequivalent ways via estimations of the escape rate \cite{kramers,melnikov} (and the references therein) on the one hand, and the first passage time \cite{ornigotti} (and the references therein), on the other.
Physically, the first method considers  possible returns of the Brownian particle into the well \cite{melnikov} that includes the semiclassical treatment by using the Wigner function master equation \cite{coffey1,coffey2,coffey3,coffey4}, while the first passage time regards when the
variable of interest attains for the first time a threshold value without return to the well \cite{dubkov,spagnolo1,spagnolo2,spagnolo3,burada,ornigotti,valenti,kramers,hangi,masoliver,kutnera,srividya}.

Investigation of the rotation stability in this paper, by utilizing the Caldeira-Leggett master equation as emphasized above, is two-fold.
First, we introduce and use a quantum-mechanical counterpart of   the first-passage time (FPT) method. Second, we use the standard method \cite{JJD1} of quantifying stability by the standard deviations of the angle and  angular momentum observables. Predictions of the two methods are mutually consistent, e.g. \cite{kutnera}: the smaller the FPT the faster (and larger) the increase of the related standard deviation.

Those methods require the knowledge of the first and the second moments of the relevant observables.
With the use of the Caldeira-Leggett master equation, we derive analytical expressions for the differential equations for the moments of both the angle of rotation $\hat\varphi$ and the angular momentum $\hat L_z$ quantum observables. We obtain an {\it infinite set} of the coupled first-order differential equations for the moments. Solving an infinite set of equations is an open, poorly-solved mathematical problem even for the commutative variables \cite{mao,frewer,kuehn}. To this end, different methods are used for obtaining the approximate/plausible solutions that are considered on the case-to-case basis. Our solution to this problem is perturbative, without the use of the standard quantum-mechanical perturbation methods. That is, we assume the small real constant $b$. Then, neglecting the terms of the order of $b^2$, we obtain a closed set of  differential equations.

Solutions to the differential equations are found partially in the analytic form. An extensive quantitative analysis of the solutions is performed for the different parameter ranges and  dynamical regimes.  The findings reveal a physically rich behavior that does not provide  simple recipes or straightforward protocols for utilizing the rotator stability. Rather, the combinations of the different stability criteria should be separately considered in order to provide {\it optimal} conditions for the rotator dynamics.

The structure of this paper is as follows: In Section II we present  details regarding the physical model. The general methodological details that include description of our method for obtaining the closed set(s) of the differential equations for the moments of the angle and  angular momentum observables are presented in Section III; in Appendix we provide the complete matrix for the system of the differential equations for the moments up to the fourth order. In Section IV we introduce and investigate a quantum-mechanical counterpart of the first passage time, that we dub quantum first-passage-time (QFPT), for the angle-observable. An emphasis is placed on the numerical investigation of the QFPT-dependence on the number of blades of the rotator. In Section V, we investigate dynamics of the standard deviations of both observables. Extensive analysis of the parameter dependence provides a rather rich physical findings that are briefly commented on  in the respective sections IV.C and V.C. On this basis, a discussion of the obtained results and the general remarks on the task of  practical utilizing  the  rotator stability are presented in Section VI. Those remarks distinguish the role of both the small parameter $b$ as well as of the propeller's size $N$. Section VII is our conclusion.

\section{The model assumptions}

In this paper we adopt the general model and terminology regarding the propeller-shaped molecular rotators \cite{kottas} (and the references therein) that use  analogy with the macroscopic counterparts,
which consist of  mutually independent "blades" and their independent local environments. Therefore, the total moment of inertia of the rotator is the sum of  individual-blades
momentums of inertia. Analogously, for  independent local environments, e.g. in the
scattering-model of interaction of the blades with the environmental molecules, the total
strength of interaction with the environment can be modeled as a sum of the strengths of
interaction for the individual blades. Thus linear dependence on the number $N$ of the blades
 follows for both the moment of inertia as well as for the total damping factor \cite{JJD1}; cf. Figure 1 in Ref. \cite{JJD1}.
This presents a limitation of our considerations: for the mutually dependent blades and/or mutually dependent local environments,
the linear dependence may be expected to be lost.

We consider a weakly cubic potential for the rotator of the form of $V(\varphi) = I\omega^2\varphi^2/2 + b \varphi^3$, where $\vert b\vert$
is a small parameter. Quantization of the angle and  angular momentum observables is a subtle task \cite{breitenberger,deck}. Particularly, dealing with the finite rotations or with the $2\pi$ rotational symmetry of the model Hamiltonian requires specific quantization procedure, cf. e.g. \cite{roulet}. Hence
direct quantization of the angle variable adopted in this paper, symbolically $\varphi \to \hat\varphi$,  allows {\it only the small rotations} to be considered.
Then the finite rotations can be realized only by a (finite) set of  small consecutive rotations.

Realistic rotators are assumed to be placed in some external fields and/or resting on a solid surface thus producing effective external field for  rotation. Typically,
such scenarios can be modeled by a weakly cubic potential.
In such situations, the free choice of the external field can introduce the $N$-independent parameters $\omega$ and/or $b$. On the other hand, the intrinsic potential for rotation \cite{kottas} introduces
dependence of the rotator's energy on the number of blades. In some cases, the number $N$ of the blades determines the number $n$ of the local minimums for the potential, which is typically modeled
as the cosine function of the form $W\cos(n\varphi)/2$, with the energy-barrier height $W$ \cite{kottas}. While the details in this regard can be found in the literature, e.g. \cite{kottas} (and the references therein), our restriction to small rotations reduces the model-potential to only one (local) minimum. That is,  the assumption of  small rotations practically excludes the transitions between the minimums
and reduces the total potential to only one such minimum. Now, following the standard wisdom, a local minimum can be approximated by the quadratic potential, with the cubic term as the first approximation--which is our case of study.  It is worth emphasizing, that inclusion of the often regarded quartic and sixtic perturbations \cite{fluge}, ($\hat\varphi^n, n=4,6$), introduces the changes of the  shape of the potential far from the local minimum and therefore requires a separate analysis.

Therefore there is not a general $N$-dependence of the parameters $\omega$ and $b$. In order to compensate for this lack of the general case, we separately investigate the
cases for the different combinations of the values for both $\omega$ and $b$. To this end, it is worth emphasizing that placing $b=0$ for the calculations presented in Section V returns the results obtained for the pure harmonic model \cite{JJD1}. Thus, in order to facilitate the calculations, we formally consider the parameters $\omega$ and $b$ as  constants, whose values are independently varied. Interestingly, the results presented in the following  sections, {\it qualitatively do not change} with the variations of the values of $\omega$ and $b$.

The cubic potential is  size-dependent but of the same form for every number $N$ of the blades. In Figure 1 we depict the potential for  $N=1$  where the local maximum $\varphi_{\max}$ is emphasized.

\begin{figure*}[!ht]
\centering
    \includegraphics[width=0.4\textwidth]{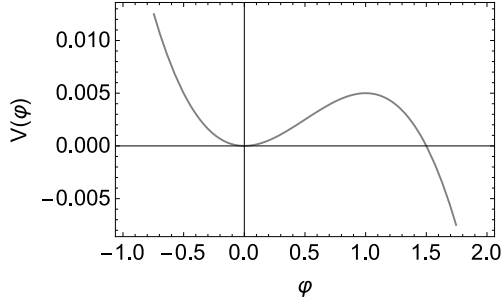}
\caption{The cubic potential for the choice of the parameters: $\omega = 0.1, I_{\circ} = 3, b=-0.01$.}
\end{figure*}

For every $N$, $1.5\varphi_{\max}^{(N)}$ lies on the horizontal axis, while $1.6\varphi_{\max}^{(N)}$ lies below the horizontal axis in Fig.1.
The choice of $\omega=0.1$ and $b=-0.01$ is used to present the results in Section IV, while the choice $\omega=1$ and $b=-0.001$ is used to present the results in Section V.

\section{The task: dynamical stability of the molecular propellers}
\label{druga}

The molecular propellers are recognized as the main candidates for the realistic artificial nanoscale cogwheels; their study is currently rather extensive so we emphasize just a small sample of the existing literature, e.g. \cite{kottas,goodsell,kudernec,balzani,hutchison} (and the references therein).
While in principle there is no an estimate of the limitation on the number $N$ of the blades yet, our considerations are restricted to the maximum $N \le 10$ \cite{JJD1}.
The physical units $I_{\circ}$ and $\gamma_{\circ}$  stand for the moment of inertia and the damping factor that  regard the molecules of the different chemical species and geometry--the only model assumption  in considering the size of the propeller is the linear scaling, $I=N I_{\circ}$ and $\gamma=N \gamma_{\circ}$, for the molecule moment of inertia and the  damping factor, respectively \cite{JJD1}.

For the one-dimensional (planar) rigid rotator, the Hamiltonian reads:

\begin{equation}
\hat H = {\hat L_z^2 \over 2I} + \hat V,
\end{equation}

\noindent where the cubic potential reads: $\hat V = I\omega^2 \hat\varphi^2/2 + b \hat\varphi^3$ for the rotator of the circular frequency $\omega$.

We utilize the standard Caldeira-Leggett master equation \cite{breuer,CL}:

\begin{equation}
{d \hat\rho(t)\over dt} = -{\imath\over\hbar} [\hat H, \hat\rho(t)] -
{\imath\gamma\over\hbar} [\hat \varphi, \{\hat L_z, \hat\rho(t)\}] - {2I\gamma
k_B T\over \hbar^2}[\hat \varphi, [\hat \varphi, \hat\rho(t)]].
\end{equation}

\noindent as the model dynamics for the system; the curly brackets denote the anticommutator. The only degree of freedom is the rotational angle $\hat\varphi$ with its conjugate angular momentum $\hat L_z$ (for the rotation around the $z$-axis), with the notation as defined above.

We regard eq.(2) as a ''phenomenological'' equation \cite{ferialdi}, in the sense of {\it not} imposing any restrictions on the values of the rotator's frequency $\omega$, the damping factor $\gamma_{\circ}$ and the bath's temperature $T$. Therefore our analysis incorporates the usual under- and over- damped regimes.

In the next two sections we identify and analyse the quantitative measures of the dynamical stability of the rotator described by eqs. (1) and (2). Those measures are based on the first and  second moments for the $\hat \varphi$ and  $\hat L_z$ observables. After a simple algebra, the linear differential equations for the moments follow from eq.(2) in the general form:

\begin{equation}
{d\langle \hat A\rangle \over dt} = {-\imath \over \hbar} \langle [\hat A, \hat H]\rangle + {\imath\gamma\over \hbar} \langle \{\hat L_z, [\hat\varphi, \hat A] \}\rangle - {2I\gamma k_BT\over \hbar^2} \langle[\hat\varphi, [\hat\varphi, \hat A]] \rangle,
\end{equation}

\noindent where $\langle\ast\rangle = tr(\ast \hat\rho)$.

Hence the  set of the coupled first-order differential equations that can be presented in the matrix form:

\begin{equation}
{d\over dt} X = \mathcal{M} X + K,
\end{equation}

\noindent with the vector $X$ composed of the moments of the form $\langle \hat \varphi^m\rangle$, $\langle \hat L_z^n\rangle$ and $\langle \hat\varphi^m \hat L_z^n + \hat L_z^n \hat \varphi^m \rangle$, and the $K$ vector collecting the inhomogeneous part of the set of the differential equations. The general solution of eq.(4) can be written in the integral form of:

\begin{equation}
X(t) = \exp(\mathcal{M}t) X(0)+ \exp(\mathcal{M}t) \int_0^t ds \exp(-\mathcal{M}s) K(s),
\end{equation}

\noindent which is particularly suited for the finite-rank matrix $\mathcal{M}$.

In Appendix, we provide the data for equations (4) and (5) for the moments up to the fourth order, where it is obvious that the set of the coupled differential equations is not closed. In order to overcome this problem \cite{mao,frewer,kuehn}, we introduce the following, perturbation-like procedure.

For every moment denoted $A_i(b,t)\equiv \langle \hat A_i(b,t)\rangle$, we look for the approximate solution for small positive $b$ in the form:

\begin{equation}
A_i(b,t) = A_i(b=0,t) + f_i(b,t) = A_i(b=0, t) + b f_i^{(1)}(t)+ b^2  f_i^{(2)}(t)+...
\end{equation}

\noindent Therefore, the knowledge of the ''unperturbed'' $A_i(b=0,t)$ reduces our task to solving the set of the coupled differential equations for the $f_i$s, such that
$f_i=0, \forall{i}$ for $b=0$.

From Appendix follow the exact differential equations  for the first and second moments:

\begin{eqnarray}
&\nonumber&
{d\langle\hat\varphi\rangle \over dt} = {1\over I} \langle \hat L_z\rangle,
\\&&\nonumber
{d\langle\hat L_z\rangle \over dt} = - I\omega^2 \langle \hat\varphi\rangle  - 2\gamma \langle \hat L_z\rangle - 3b \langle \hat\varphi^2\rangle,
\\&&\nonumber
{d\langle\hat\varphi^2\rangle \over dt} = {1\over I} \langle \hat\varphi \hat L_z + \hat L_z\hat\varphi\rangle,
\\&&\nonumber
{d\langle \hat\varphi \hat L_z + \hat L_z\hat\varphi\rangle\over dt} = -2I\omega^2 \langle \hat\varphi^2\rangle - 2\gamma \langle \hat\varphi \hat L_z + \hat L_z\hat\varphi\rangle + {2\over I} \langle \hat L_z^2\rangle - 6b
\langle \hat\varphi^3\rangle,
\\&&
{d\langle\hat L_z^2\rangle \over dt} =  -I\omega^2 \langle \hat\varphi \hat L_z + \hat L_z\hat\varphi\rangle - 4\gamma \langle \hat L_z^2\rangle  - 3b \langle \hat\varphi^2 \hat L_z + \hat L_z\hat\varphi^2\rangle + 4I\gamma k_BT.
\end{eqnarray}

Substituting eq.(6) into eq.(7) while keeping only the terms linear in the constant $b$ (that is, while neglecting the terms of the form $b f_i$), we obtain the following set of the differential equations for the first corrections:

\begin{eqnarray}
&\nonumber&
{d f_1^{(1)} \over dt} = {1\over I} f_2^{(1)},
\\&&\nonumber
{d f_2^{(1)} \over dt} = - I\omega^2 f_1^{(1)}  - 2\gamma f_2^{(1)} - 3 \langle \hat\varphi^2\rangle_{b=0},
\\&&\nonumber
{d f_3^{(1)} \over dt} = {1\over I} f_4^{(1)},
\\&&\nonumber
{d f_4^{(1)}\over dt} = -2I\omega^2 f_3^{(1)} - 2\gamma f_4^{(1)}  + {2\over I} f_5^{(1)} - 6 \langle \hat\varphi^3\rangle_{b=0},
\\&&
{df_5^{(1)}\over dt} =  -I\omega^2 f_4^{(1)} - 4\gamma f_5^{(1)}  - 3 \langle \hat\varphi^2 \hat L_z + \hat L_z\hat\varphi^2\rangle_{b=0},
\end{eqnarray}

\noindent
The ''unperturbed'' moments indexed in eq.(8) by "$b=0$" follow from the closed sets of the differential equations for the case $b=0$, cf. Appendix.
In eq.(8) we can recognize the inhomogeneous part presented in the vector form:  $P^T=\{0, -3\langle \hat\varphi^2\rangle_{b=0}, 0, -6 \langle \hat\varphi^3\rangle_{b=0},  - 3 \langle \hat\varphi^2 \hat L_z + \hat L_z\hat\varphi^2\rangle_{b=0} \}$; the superscript "T" denotes the operation of  the matrix transposition. Therefore, as desired, eq. (8) represents a {\it closed} set of equations for the corrections $f_i$s and hence also for the first and  second moments in eq.(7).
Actually, there appear  two independent closed sets of equations for the first and the second moments to be separately analyzed in the next two sections.

The general solution of eq.(8) is of the form of eq.(5):

\begin{equation}
F(t) = \exp(\mathcal{\mu}t) F(0)+ \exp(\mathcal{\mu}t) \int_0^t ds \exp(-\mathcal{\mu}s) P(s),
\end{equation}

\noindent where the vector $X^T$ is replaced by the vector $F^T= \{f_1, f_2, f_3, f_4, f_5\}$, while the matrix $\mu$ follows from eq.(8):

\begin{equation}
\mu = \left({
  \begin{array}{cccccccccccccc}
    0       & 1/I & 0 & 0 & 0   \\
    -I\omega^2       & -2\gamma &   0 &  0 & 0  \\
    0 &0 &  0 & 1/I &  0 \\
    0 & 0 & -2I\omega^2 & -\gamma2 &   2/I    \\
0 & 0 & 0 & -I\omega^2 &-4\gamma
  \end{array}}
\right)
\end{equation}

Hence the approximate solutions for the first-order corrections for the first and  second moments, while $f_i=b f_i^{(1)}$, read:

\begin{eqnarray}
&\nonumber&
\langle\hat\varphi\rangle  = \langle\hat\varphi\rangle_{b=0}+  f_1,
\\&&\nonumber
\langle\hat L_z\rangle  = \langle\hat L_z\rangle_{b=0} +  f_2,
\\&&\nonumber
\langle \hat\varphi^2\rangle = \langle \hat\varphi^2\rangle_{b=0} + f_3,
\\&&
\langle\hat L_z^2\rangle  = \langle\hat L_z^2\rangle_{b=0} +  f_5,
\end{eqnarray}

\noindent whence the solutions for the standard deviations $\Delta \hat\varphi$ and $\Delta \hat L_z$ readily follow:

\begin{eqnarray}
&\nonumber&
\Delta \hat\varphi  = \sqrt{\langle \hat\varphi^2\rangle_{b=0} +  f_3 - \left(\langle\hat\varphi\rangle_{b=0}+  f_1\right)^2},
\\&&
\Delta \hat L_z  = \sqrt{\langle\hat L_z^2\rangle_{b=0} +  f_5 - \left(\langle\hat L_z\rangle_{b=0} +  f_2\right)^2}.
\end{eqnarray}

\noindent For different orders of approximation,  different solutions of the corrections $f_i$s are obtained; the increase in the order of approximation increases the rank of the matrix $\mu$ in eq. (9).

\section{The first passage time}
% or \section{Numerical results and discussion}
\label{treca}

The first passage time (FPT) method regards the minimum time, $t_{FPT}$, needed for a system to cross a threshold value for the variable of interest. The shorter $t_{FPT}$ the faster the transition from the initial state and hence the less stable the system. The method has a long tradition in physics, engineering, and natural sciences and has recently been used for describing  financial market volatility \cite{valenti,masoliver}.

In the classical physics context, the so-called mean FPT is of interest that is defined as the arithmetic mean of the FPTs for different, stochastically chosen (numerically: sampled) trajectories.
Quantum-mechanical counterpart of the (mean-)FPT is an ill-defined and the context-sensitive concept \cite{kumar,pawela,tao,grot}. Since it is linked with the deterministic classical trajectories, there is not a straightforward quantum mechanical definition. In general, quantum models may even end up with the non-positive probability density \cite{kumar}. Approaching the classical meaning may call for the intermediate quantum measurements \cite{kumar,pawela}, while other definitions regard e.g.  dynamics of the system where the FPT is linked with the system's state transition \cite{tao}--very much like the general task of the time bound for the quantum state change \cite{grot,margolus,DC}.

In this paper we introduce a quantum mechanical counterpart, which we dub "quantum FPT" (QFPT), by investigating the minimum time needed for the first moment $\varphi = \langle \hat\varphi\rangle$ to take some threshold value $\varphi_{th}$ for the chosen initial $\varphi_{\circ}$ value; further comments on this can be found in Discussion section. Bearing in mind the constraint of our considerations, i.e. the small allowed rotations (i.e. a finite set of  small rotations), we consider $\vert\varphi_{th} - \varphi_{\circ}\vert \approx 10^{-4}$. That is, for every chosen initial $\varphi_{\circ}$, we assume a close threshold $\varphi_{th}$ value for the angle observable, and numerically calculate $t_{QFPT}$ as the minimum time needed for the transition $\varphi_{\circ} \to \varphi_{th}$.

Dependence of $t_{QFPT}$ on the damping factor $\gamma$ and the bath's temperature $T$ is widely investigated.  Our main goal in this section is to extend the standard analysis  by investigating the role of the propeller {\it size}.

Comparison of  different sizes of the propeller rotators is performed by comparing  numerically obtained values for $t_{QFPT}$ for the chosen initial positions, $\varphi_{\circ}^{(N)} \in \{1.1\varphi_{\max}^{(N)}, 1.3\varphi_{\max}^{(N)}, 1.5\varphi_{\max}^{(N)}, 1.6\varphi_{\max}^{(N)}\}$, for the number $N \in \{1, 2, 3,..., 10\}$ of the blades.
That is, our goal in this section is to obtain analytical expression for the $t_{QFPT}$ dependence on the number $N$ of the blades, denoted $t_{QFPT}(N)$.

With the aid of eq.(A.4) in the Appendix section for the standard deviation $\Delta\hat\varphi$, the first pair of equations (for the first moments) in eq.(8) is straightforward analytically to solve. Due to eq.(11), the corrections $f_i$ contribute to the initial values of the first moments by introducing additional size-dependence for $\langle\hat\varphi\rangle$. Therefore, in order to facilitate comparison of the results for the propellers of  different sizes, we choose the initial values for the corrections: $f_i^{(1)}(t=0) = 0, i=1,2,...,5$. Then, due to eq.(9), follow the analytical expressions for the corrections:

\begin{eqnarray}
&\nonumber&
f_1  =  3b
\left(-{1\over I\omega^2} +
{ e^{-\gamma t}\over I\omega^2}
 (\cosh(\Omega t) + {\gamma\over  \Omega} \sinh(\Omega t)
\right)
\times
\\&&\nonumber
\bigg({k_BT\over I\omega^2} + {e^{-2\gamma t}(B^2+q) \sinh^2(\Omega t)\over I^2\Omega^2} +\\&&\nonumber
{  e^{-2\gamma t}(2AB+r)   (2\gamma \sinh^2(\Omega t) +\Omega \sinh(2\Omega t))
\over 2I\Omega^2} +\\&&\nonumber
{e^{-2\gamma t} (A^2+p) (-\omega^2 \cosh^2(\Omega t) + \gamma^2 \cosh(2\Omega t) + \gamma\Omega \sinh(2\Omega t))
\over \Omega^2} +\\&&
{k_BT e^{-2\gamma t} (\omega^2 - \gamma^2\cosh(2\Omega t) - \gamma\Omega \sinh(2\Omega t))
\over I\omega^2\Omega^2}\bigg)
\end{eqnarray}

\noindent  and

\begin{eqnarray}
&\nonumber&
f_2  = -
{3be^{-3\gamma t} \over 2I^2 \omega^2 \Omega^3} \sinh(\Omega t) \times
\\&&\nonumber
\bigg(2\omega^2(B^2+q)\sinh^2(\Omega t)  + \\&&\nonumber
2I^2\omega^2(A^2+p)  (-\omega^2 \cosh^2(\Omega t) + \gamma^2 \cosh(2\Omega t) + \gamma\Omega \sinh(2\Omega t)) +\\&&\nonumber
I \Big(
2k_BT\gamma^2 (e^{2\gamma t} - \cosh(2\Omega t)) +
\\&&\nonumber
2\gamma (\omega^2 (2AB+r) \sinh^2(\Omega t) -  k_BT\Omega\sinh(2\Omega t))
-
\\&&
\omega^2 (2k_BT(e^{2\gamma t}-1) - \Omega(2AB+r)\sinh(2\Omega t))
\Big)
\bigg).
\end{eqnarray}

\noindent from which it is obvious that the initial  condition $f_i^{(1)}(t=0)=0$ is satisfied for both $i=1,2$, while the asymptotic expressions read: $\lim_{t\to\infty} f_1^{(1)} = -3k_BT/(I\omega^2)^2$ and $\lim_{t\to\infty} f_2^{(1)}=0$. The constants appearing in eqs.(13) and (14) are as follows: $A=\langle\hat\varphi(0)\rangle$, $B=\langle\hat L_z(0)\rangle$, $p=(\Delta\hat\varphi(0))^2$, $q=(\Delta\hat L_z(0))^2$,
$r=\sigma_{\varphi L}(0)$. This, somewhat cumbersome notation, is used in order to facilitate the presentation of the quantum Cauchy-Schwarz inequality, $\sigma_{\varphi L} \equiv \langle\hat\varphi \hat L_z +\hat L_z\hat\varphi \rangle - 2\langle\hat\varphi\rangle \langle\hat L_z\rangle \le 2 \Delta\hat\varphi \Delta\hat L_z$, which is satisfied by the choice of the initial values, $p=0.01, q=0.005, r=0$, while $A=\varphi_{\circ}^{(N)}$ and $B=1.2$.

Substituting $\langle\hat\varphi\rangle_{b=0}$ from eq.(A.3) in Appendix and eq.(13) into the first equation in eq.(11), we obtain the general form of the first-order solution for $\langle\hat\varphi(t)\rangle$.
Then we perform numerical calculation of $t_{QFPT}$ as described above for all combinations of the number $N
\in\{1,2,3,..., 10\}$ and certain values of $\gamma_{\circ}$ and $k_BT$, and the above distinguished initial angles $\varphi_{\circ}^{(N)}$ with the threshold values defined as $\vert \varphi_{th}^{(N)} - \varphi_{\circ}^{(N)}\vert \approx 0.0001, \forall{N}$.

We search for the minimum time for which the value $\varphi(t)$ attains the threshold value, $\varphi_{th}$; the equality $\varphi(t_{QFPT}) = \varphi_{th}$ (i.e. $\delta\varphi=0$ for $t=t_{QFPT}$) is presented in this section by the constant (horizontal) plane.
Different combinations of the values for $\omega$ and $b$ have been investigated. Without loss of generality, below, we provide the results for $\omega=0.1, b=-0.01$ that {\it qualitatively} present the findings for all the considered values of the parameters $\omega$ and $b$.

\subsection{The case $\gamma_{\circ} > \omega$}

We choose the values for $\gamma_{\circ}\in [0.11, 20]$ that are all larger than the chosen $\omega = 0.1$. We find only weak dependence on the initial position and practically negligible contribution of $k_BT$. There is an increase of $t_{QFPT}$ with the number $N$ of the blades, with the faster increase for smaller $N$. That is,   larger rotators are more stable.

Without  loss of generality, Figure 2 illustrates two cases of relatively small, and relatively large values of the damping factor  for different initial positions and different temperatures.

\begin{figure*}[!ht]
\centering
    \includegraphics[width=0.3\textwidth]{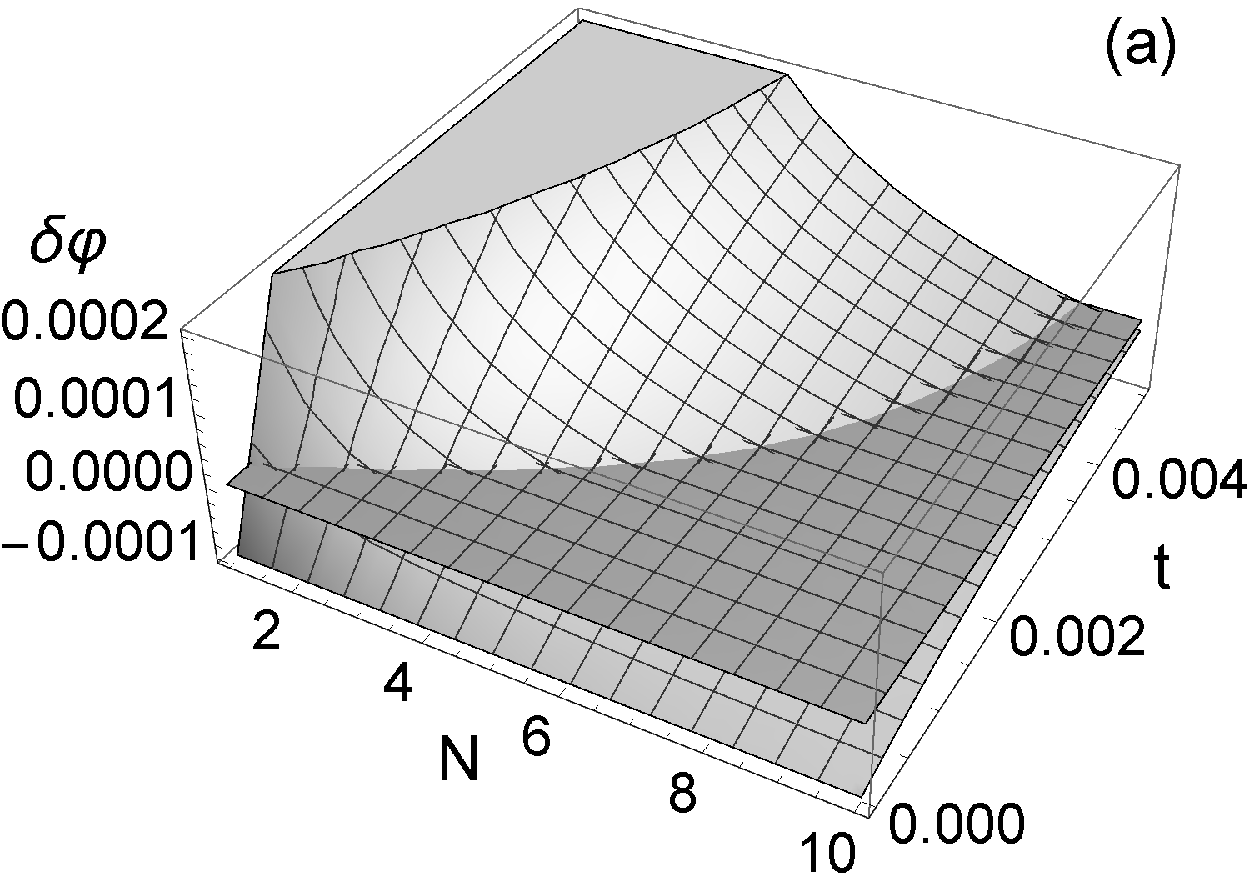}
    \includegraphics[width=0.3\textwidth]{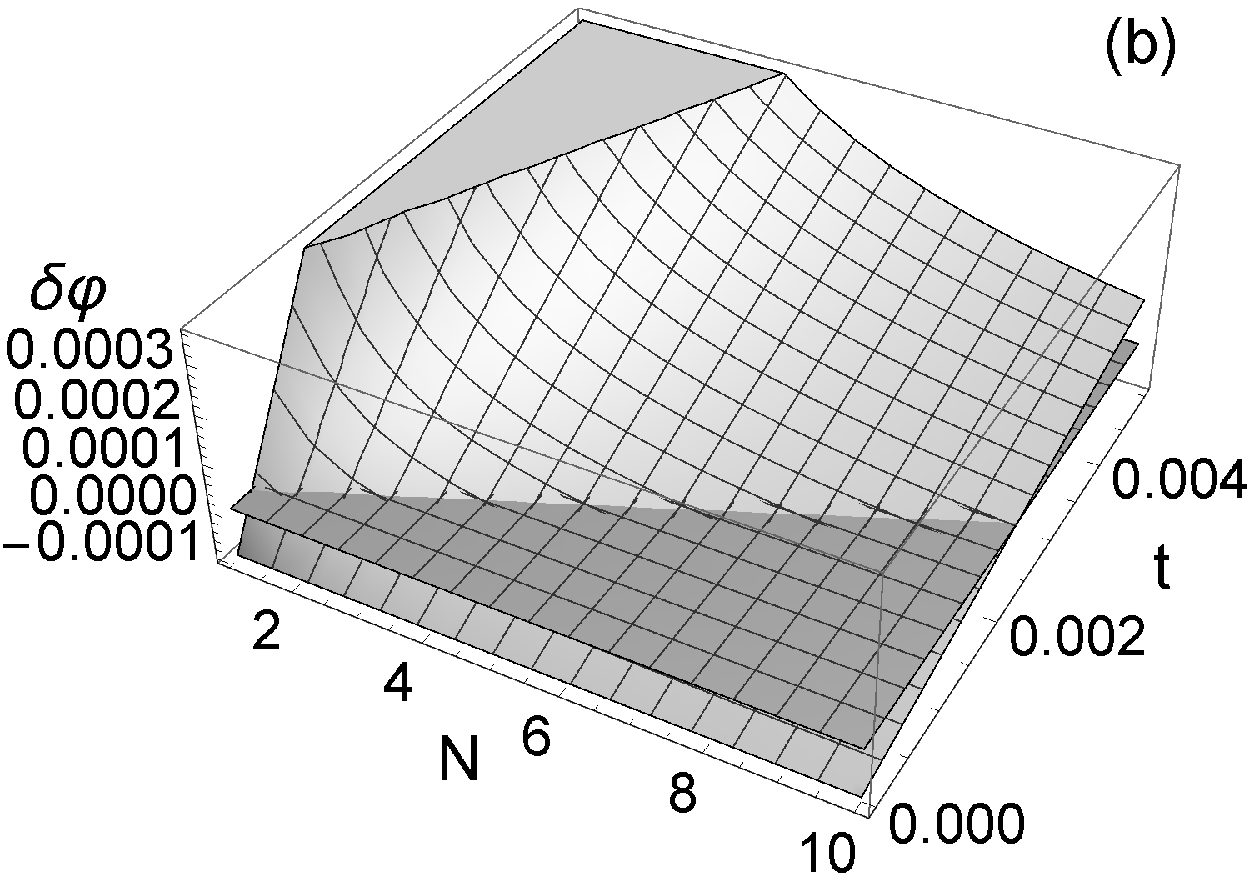}
    \includegraphics[width=0.3\textwidth]{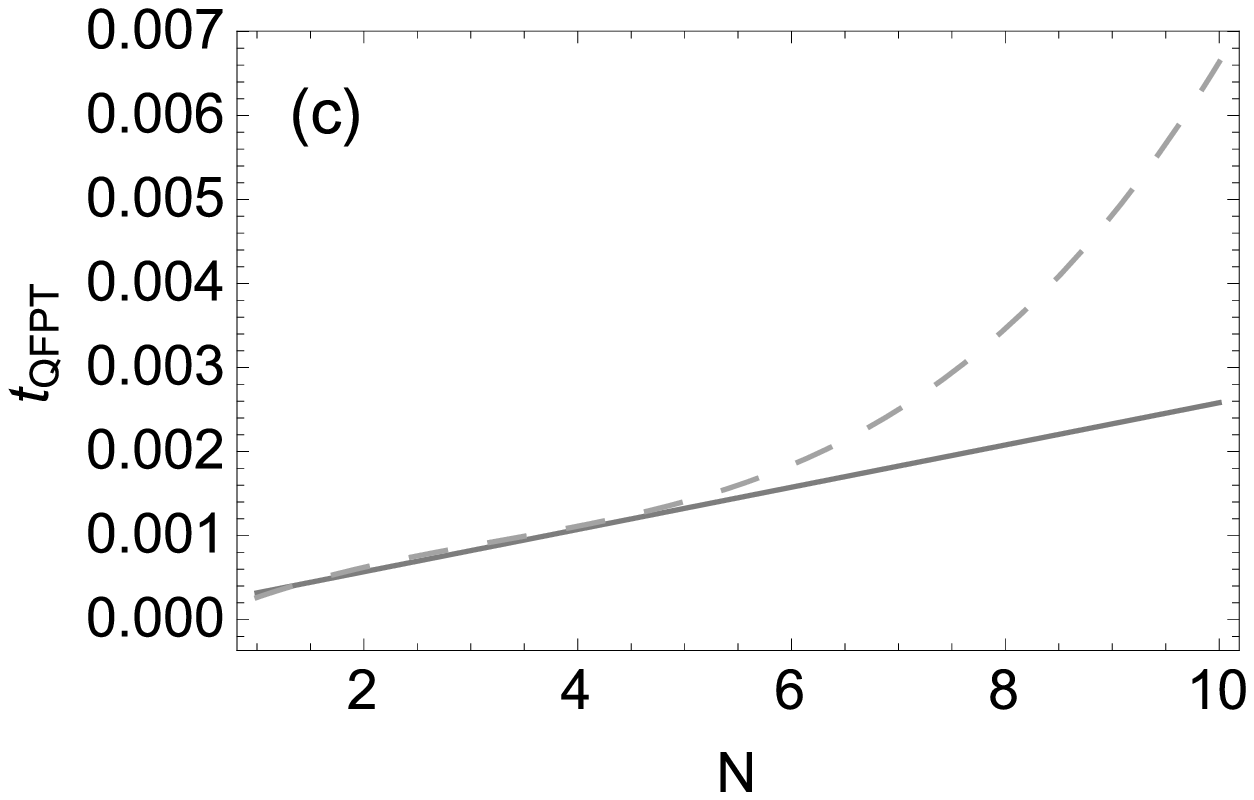}
\caption{Plot of $\delta\varphi\equiv\langle\hat\varphi(t,N)\rangle - \varphi_{th}$. Intersection with the horizontal plane determines the $t_{QFPT}$ for the choice of the parameters: (a) $\gamma_{\circ}=20, k_BT=0.001, \varphi_{\circ}^{(N)}=1.1$ and (b) $\gamma_{\circ}=0.11, k_BT=100, \varphi_{\circ}^{(N)}=1.3$. Figure (c) presents the numerically obtained dependence of $t_{QFPT}(N)$ for the (a) and (b) plots, the dashed line and the solid line, respectively.}
\end{figure*}

Figure 2(c) emphasizes the (numerically obtained) $t_{QFPT}$-dependence on the number of blades $N$. Approximate analytical expressions are [scaled yet]: (a) $-2.8+6.7 N - 1.3 N^2+0.1 N^3$, and (b) $0.67+2.5 N$. It is worth repeating: the observed patterns for $t_{QFPT}(N)$ do not change with the variation of the initial position or of the temperature $T$. The magnitude of  change of $\langle\hat \varphi\rangle$  is of the same order for the two cases.

The observed increase of $t_{QFPT}$ with the increase of the number $N$   may seem intuitively expected--the larger the system, the larger its inertia and therefore the slower the system's dynamics. That is, one may ask if the observed behavior can be reduced to the system's inertia. Below, we demonstrate that this is not the case.

By the inertial effect, we assume the effect due to the increase of the system's inertia, while all the other system parameters remain unchanged. Therefore, in order to distinguish the inertial effects,
we remove the size-dependence everywhere except in the moment of inertia, $I=N I_{\circ}$. More formally, in the above expressions, we remove the $N$-dependence from the damping factor $\gamma$ while keeping the rest of the expressions.

In Figure 3, the results are presented for the inertial case with the same choice of  system-parameters as for the general case Figure 2. Figure 3(c) distinguishes the (numerically obtained) plots for $t_{QFPT}(N)$ with the linear analytical expressions: (a) $7.5 + 24N$, and (b) $6.7+ 25N$, the solid line and the dashed line, respectively.

\begin{figure*}[!ht]
\centering
    \includegraphics[width=0.3\textwidth]{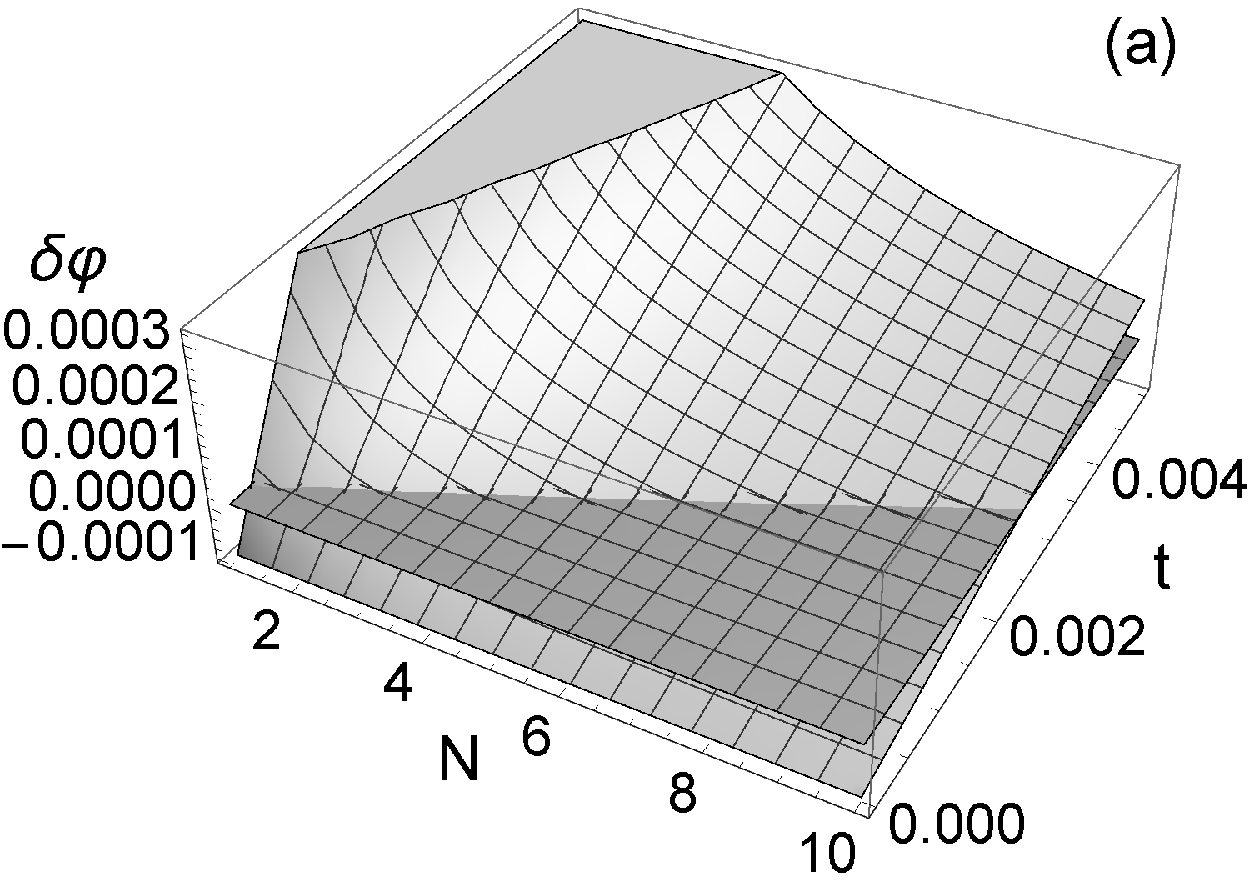}
    \includegraphics[width=0.3\textwidth]{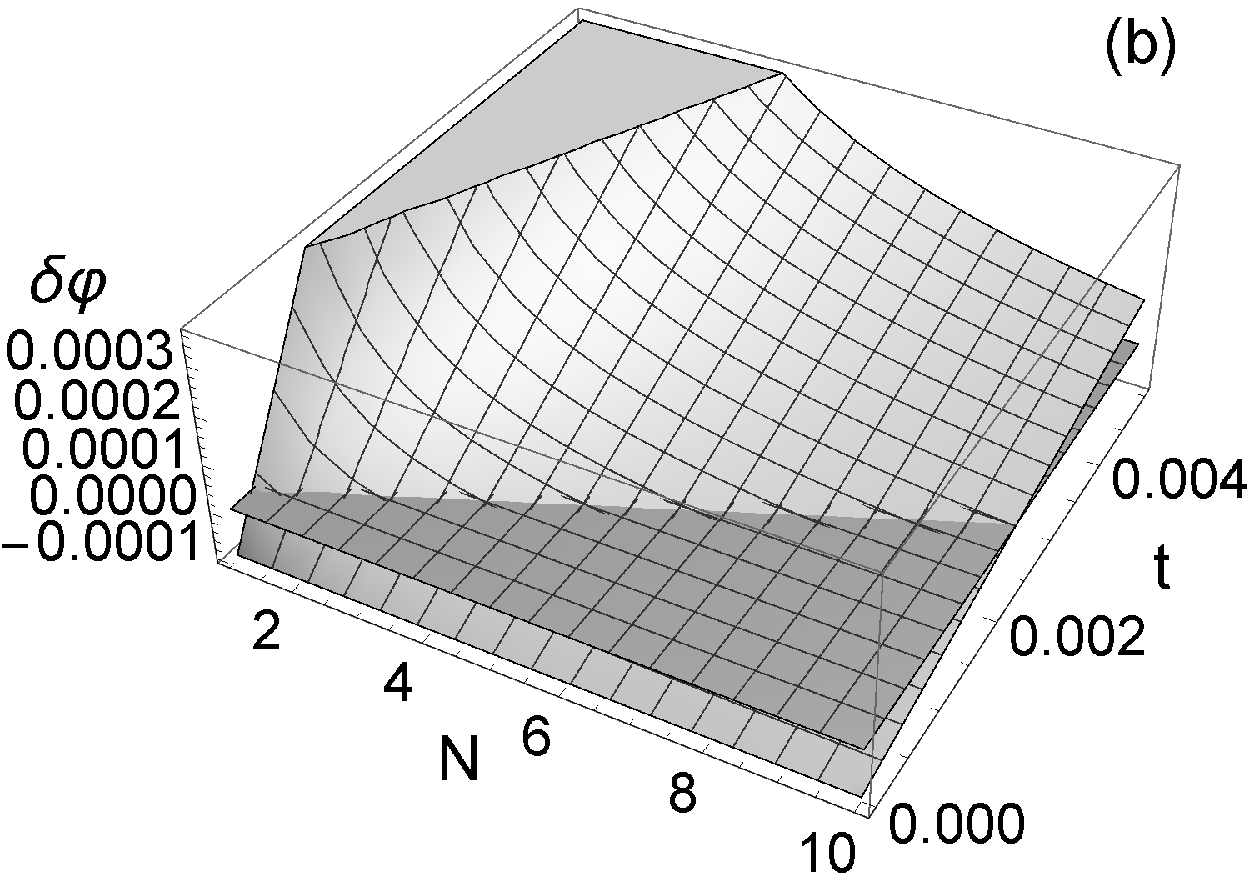}
    \includegraphics[width=0.3\textwidth]{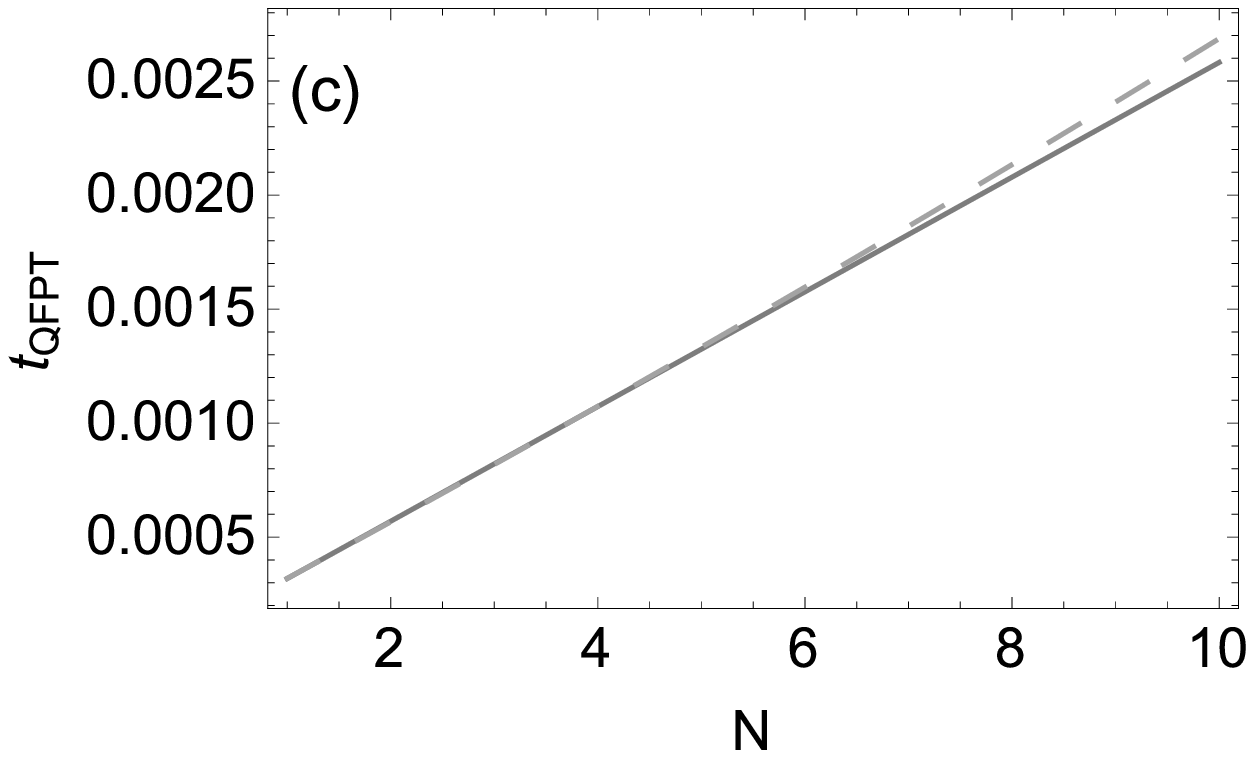}
\caption{The parameters values and the meaning of the plots is the same as for Figure 2.}
\end{figure*}

\subsection{The case $10\gamma_{\circ} < \omega$}

The choice $\gamma_{\circ}\in [0.00001, 0.0099]$ is made in order to fulfill the constraint $10\gamma_{\circ}<\omega$.
It is the general finding: the results weakly depend  on the initial position with the negligible contribution of the bath's temperature.
Therefore the same initial position and the temperature are chosen for Figure 4.

\begin{figure*}[!ht]
\centering
    \includegraphics[width=0.4\textwidth]{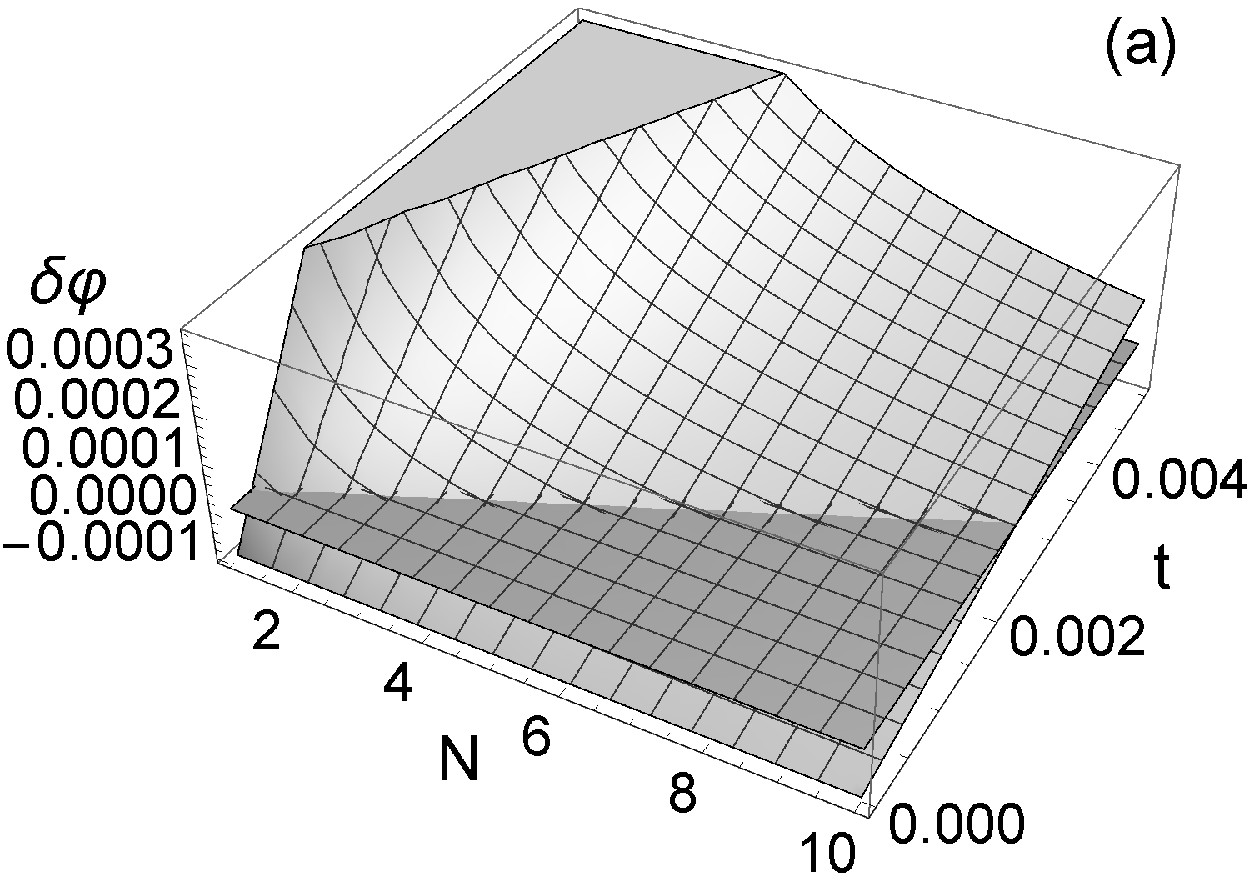}
    \includegraphics[width=0.4\textwidth]{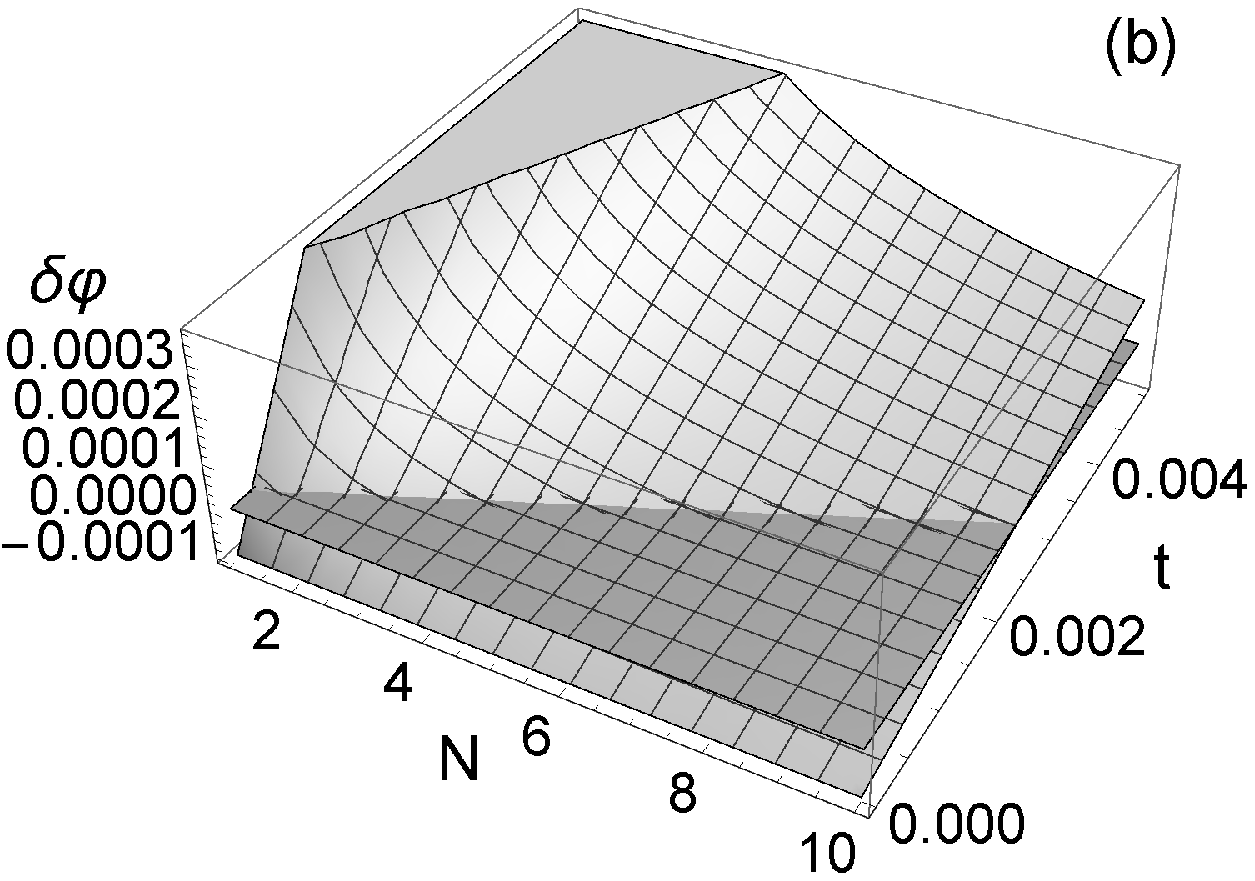}
\caption{Plot of $\delta\varphi\equiv\langle\hat\varphi(t,N)\rangle - \varphi_{th}$. Intersection with the horizontal plane determines the $t_{QFPT}$ for the choice of the parameters, $\gamma_{\circ}=0.0099$, $k_BT=0.001$ and $\varphi_{\circ}^{(N)}=1.6$, for: (a) the exact case, and (b) the inertial case.}
\end{figure*}

Figure 4 illustrates the linear dependence of $t_{QFPT}$ on $N$ that is numerically found approximately as $25.1N$ for both the exact and the inertial case.

\subsection{Comments}

The small magnitude of change of the initial $\langle\hat\varphi\rangle$ is assumed for all the considered cases. The magnitude of the change is of the order of $10^{-4}$ (or even smaller), that meets the condition of  small rotations, i.e. $\delta\varphi \ll \vert\langle\hat\varphi\rangle - \varphi_{\circ} \vert\ll  2\pi$.

The dominant factors for the investigated rotator's dynamics is the ratio $\gamma_{\circ}/\omega$. The rest of the system parameters (the initial position and the bath's temperature) is virtually of no
influence on the system's dynamics described by the first moment of the angle observable.

For the case $\gamma_{\circ} \gtrsim \omega$, the two different ''laws'' for $t_{QFPT}(N)$ are found--a weakly cubic and a linear dependence on $N$. Deviation from the approximately linear dependence is $\gamma_{\circ}$-dependent: the larger the $\gamma_{\circ}$, the smaller $N$ for which the departure becomes non-negligible. E.g., for $\gamma_{\circ}\approx 2$, the value $N\approx 7$ is found, while for $\gamma_{\circ}=20$, the value $N\approx 4$ is found.
For all other choices of $\gamma_{\circ}$ relative to $\omega$, the approximately linear
law for $t_{QFPT}(N)$ is found, including the inertial cases. The magnitude of $\delta\varphi$ is the same for all the considered combinations of the system parameters.

While for the case $10\gamma_{\circ} <\omega$ approximately the linear $t_{QFPT}(N)$ dependence is found, for larger values of $\gamma_{\circ}$, a weakly cubic dependence is found.
More precisely: the non-linear terms in $t_{QFPT}(N)$ are present for all the values of $\gamma_{\circ}$ but become quantitatively observable only for larger values of $\gamma_{\circ}$.
Therefore we conclude that the increase of $t_{QFPT}$ with the increase of the number $N$ of the propeller blades cannot be reduced to the purely inertial effect.

\section{The standard deviations}

The last three equations in eq.(8) constitute a closed system of coupled first-order differential equations while assuming the expressions are known for the third moments, $\langle \hat\varphi^3 \rangle_{b=0}, \langle \hat\varphi^2 \hat L_z + \hat L_z\hat\varphi^2\rangle_{b=0}$--which follow from eq.(A.5) in Appendix section. As it is emphasized in Appendix, analytical solutions for the third moments are rather large and not very informative. Therefore we numerically solve both eq.(A.5) and eq.(8) and present  solutions for $(\Delta\hat\varphi)^2$ and $(\Delta\hat L_z)^2$ in the graphical form with the clearly indicated dependencies on the number $N$ of the blades and on time $t$.

We use the standard fourth order Runge-Kutta method with an emphasis on the numerical stability as well as numerical reliability. Numerical stability is provided by the proper choice of the parameters so as to obtain the sufficiently large determinant for the systems of equations. Numerical reliability is additionally checked by employing the adaptive Runge-Kutta (RKF45) method for certain sensitive points discovered in the course of the computation.

The analysis has been performed for different combinations of the values of $\omega$ and $b$ with {\it qualitatively} the same results found for all the combinations. Below,  for simplicity, we present the results obtained for  $\omega=1$ and $b=-0.001$, while $k_BT\in\{0.01,0.1,1,100\}$ and $I_{\circ}=3$. For all the plots we use the initial values: $(\Delta\hat\varphi(0))^2 =0.01, (\Delta\hat L_z(0))^2=5, \sigma_{\varphi L}(0)=0$, while for the third-moments we take the same initial value $0.1$. Initial values for the first moments are: $\langle\hat\varphi(0)\rangle=1.1$, $\langle\hat L_z(0)\rangle=1.2$; as distinct from the investigation of the FPT (Section IV), those values are of the secondary importance. The only distinction between the  two regimes regards the damping factor values. For convenience, in the plots given below, we use $\sigma^2$ to denote the square of the standard deviation of the observables.

\subsection{The $\gamma_{\circ}>\omega$ case}

 For the damping factor, we choose the following values: $\gamma_{\circ}\in\{1.1,2\}$. We do not consider the larger values since the strong damping masks the investigated effects, while we do not restrict the temperature values.

\noindent {\it (a)} First, we notice the known behavior observed \cite{JJD1} for the pure harmonic oscillator: both $\Delta\hat\varphi$ and $\Delta\hat L_z$ increase with time, while $\Delta\hat\varphi$ decreases and $\Delta\hat L_z$ increases with the increase
of the number $N$ of the blades. This behavior is found for $\gamma_{\circ}=1.1$ (similarly for $\gamma_{\circ}=2$) for high temperature of $k_BT=100$, cf. Figure 5.

\begin{figure*}[!ht]
\centering
    \includegraphics[width=0.4\textwidth]{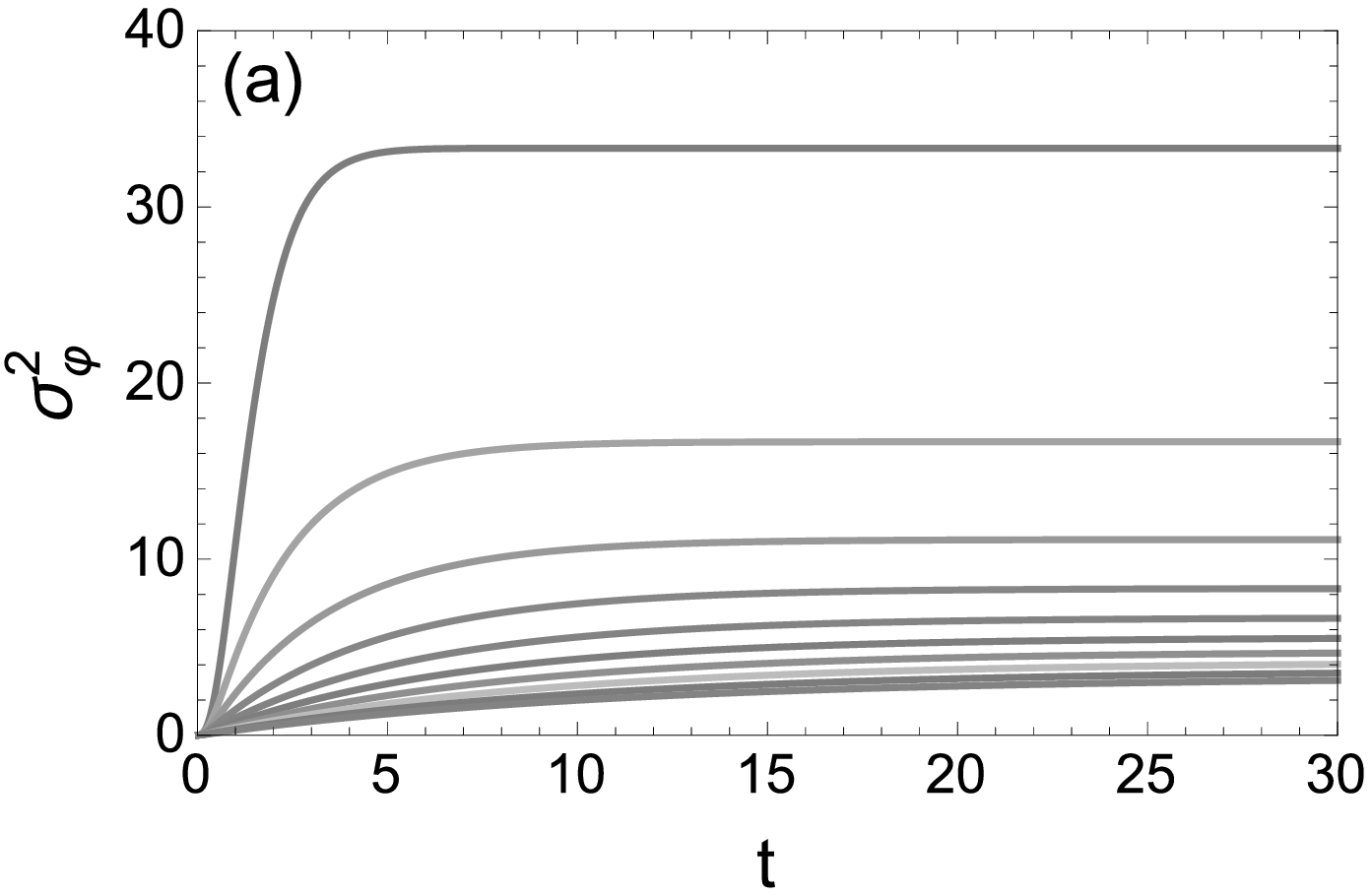}
    \includegraphics[width=0.4125\textwidth]{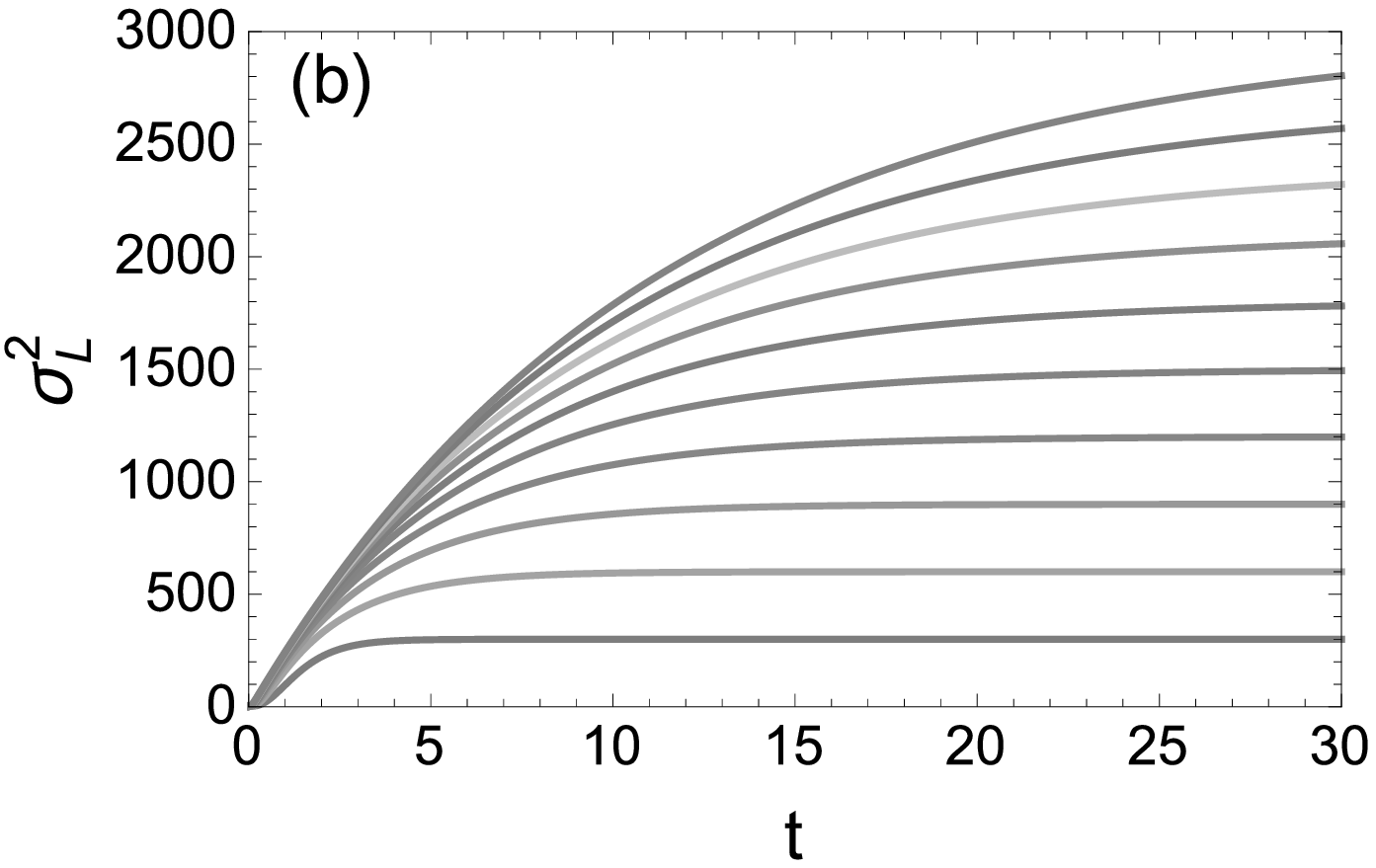}
\caption{The parameters values $\gamma_{\circ}=1.1$ and $k_BT=100$ are used for the square of the standard deviation
 for: (a) the angle (the top line for $N=1$ and the bottom line for $N=10$, consecutively) and  (b) the angular momentum (the top line for $N=10$ and the bottom line for $N=1$, consecutively).}
\end{figure*}

\noindent {\it (b)} For all other cases, a sharp initial decrease and the local maximums for short time intervals are observed for $\Delta\hat L_z$; that is followed by the increase of $\Delta \hat L_z$ with time as well as with the increase of $N$, cf. Figure 6(b). For $k_BT=0.01$ we observe  {\it decrease} of $\Delta\hat\varphi$ with the passage of time, along with the {\it existence of the minimum} $(\Delta\hat\varphi)_{min}$ for certain values of the number of  blades, around $N=3-5$, cf. Figure 6(a).

\begin{figure*}[!ht]
\centering
    \includegraphics[width=0.4\textwidth]{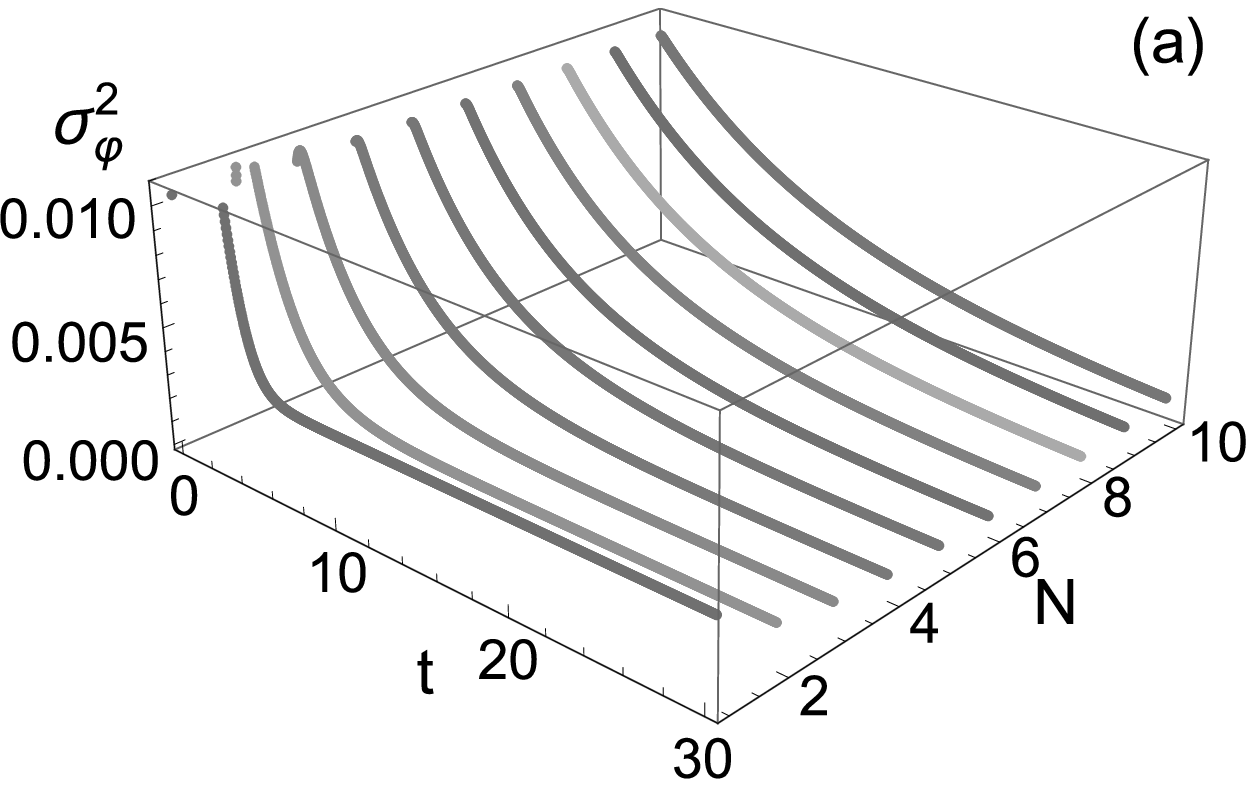}
    \includegraphics[width=0.4\textwidth]{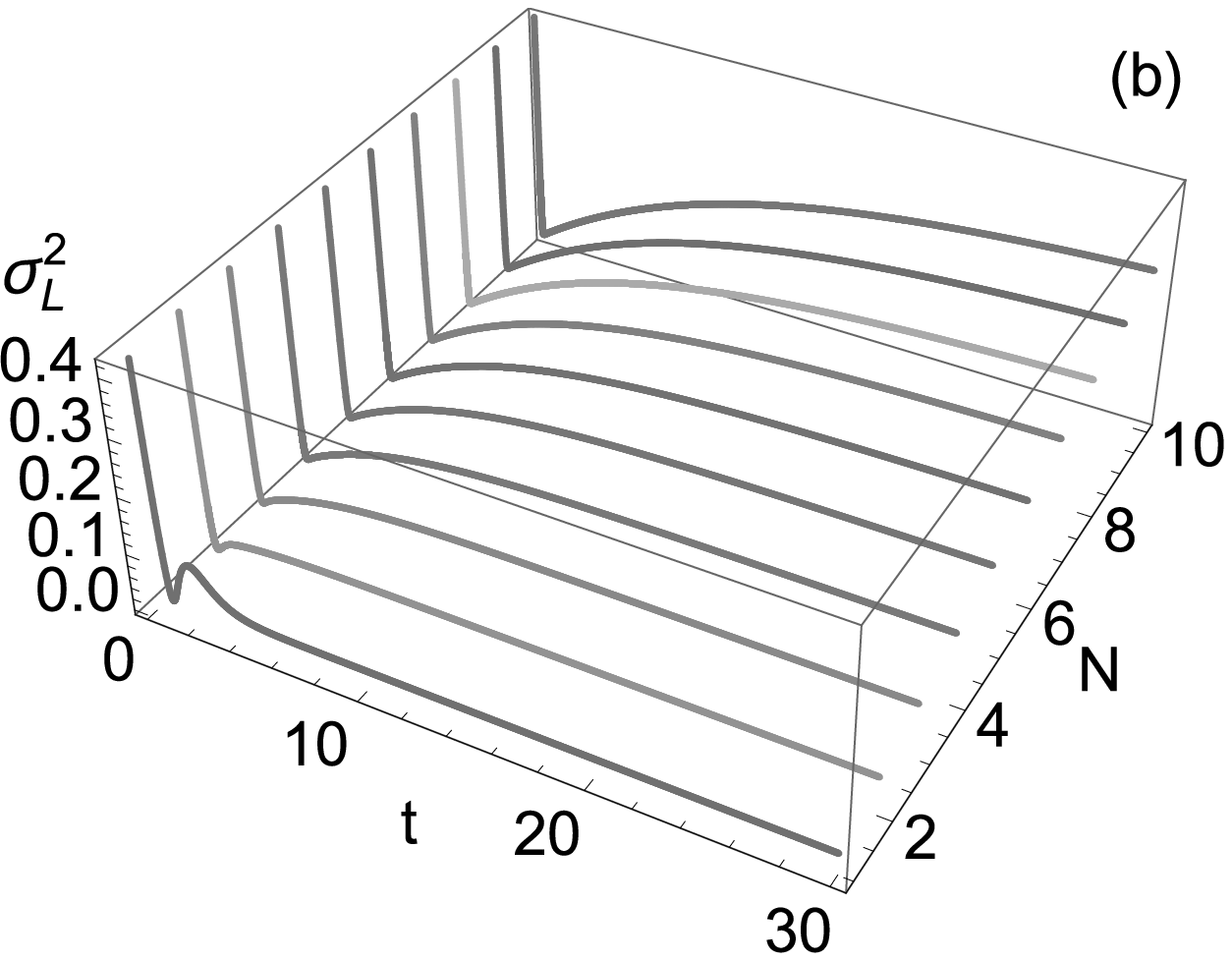}
\caption{The parameters values $\gamma_{\circ}=1.1$ and $k_BT=0.01$ are used
for the square of the standard deviation for: (a) the angle and (b) the angular momentum.}
\end{figure*}

\noindent {\it (c)} For the medium temperature, e.g. $k_BT=0.1$, in Figure 7 we observe a saturation for longer time for both $\Delta\hat\varphi$ and $\Delta\hat L_z$, that reveals a "smooth" transition between the above "extreme" cases {\it (a)} and {\it (b)}.

\begin{figure*}[!ht]
\centering
    \includegraphics[width=0.4\textwidth]{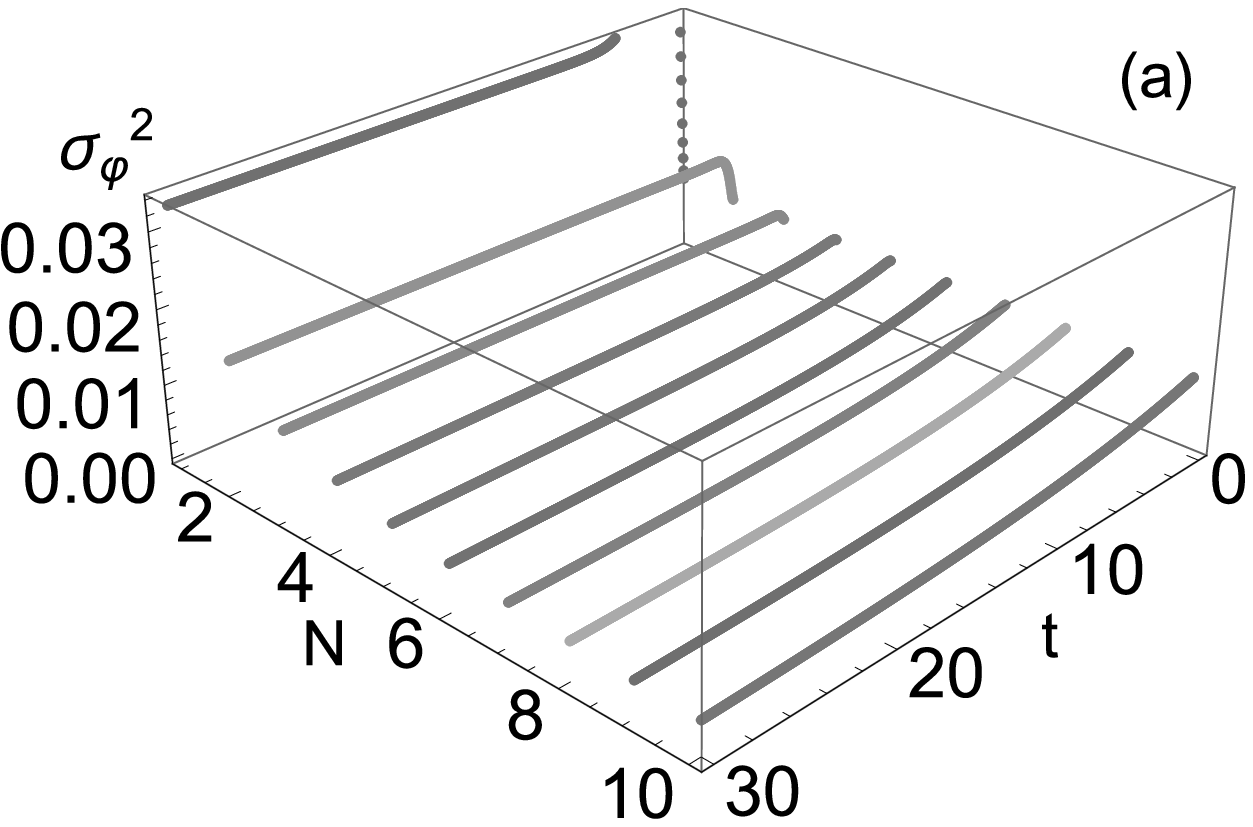}
    \includegraphics[width=0.4\textwidth]{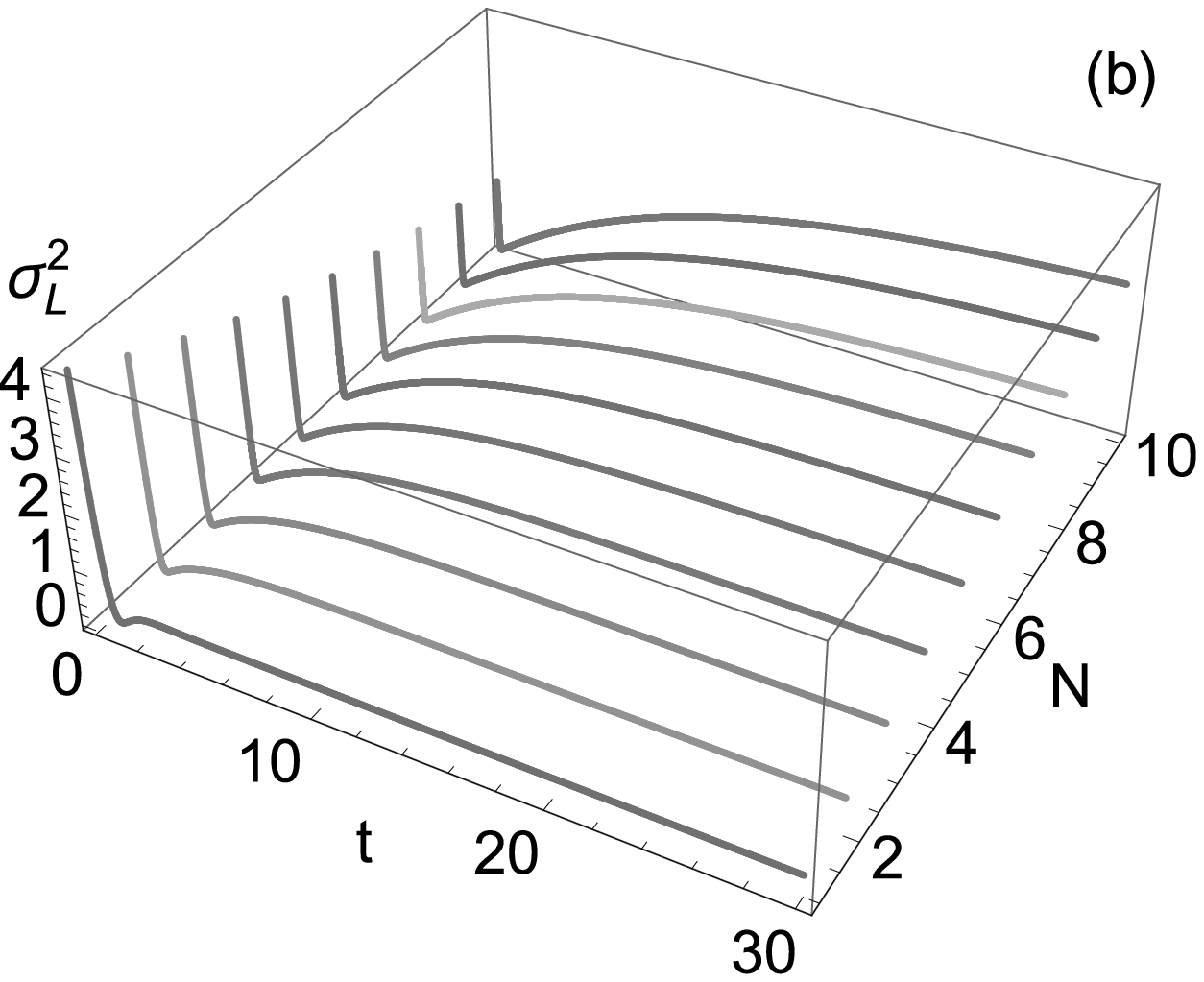}
\caption{The parameters values $\gamma_{\circ}=1.1$ and $k_BT=0.1$. are used
for the square of the standard deviation for: (a) the angle and (b) the angular momentum.}
\end{figure*}

\subsection{The $10\gamma_{\circ}<\omega$ regime}

We choose the following values for the damping factor: $\gamma_{\circ}\in\{0.00011,0.0011,0.011\}$, while we do not restrict the temperature values.

\noindent {\it (a)} Again, for $k_BT=100$, in Figure 8, we observe the standard behaviour characteristic for the pure harmonic dynamics as for the case presented in V.A{\it (a)}.

\begin{figure*}[!ht]
\centering
    \includegraphics[width=0.4\textwidth]{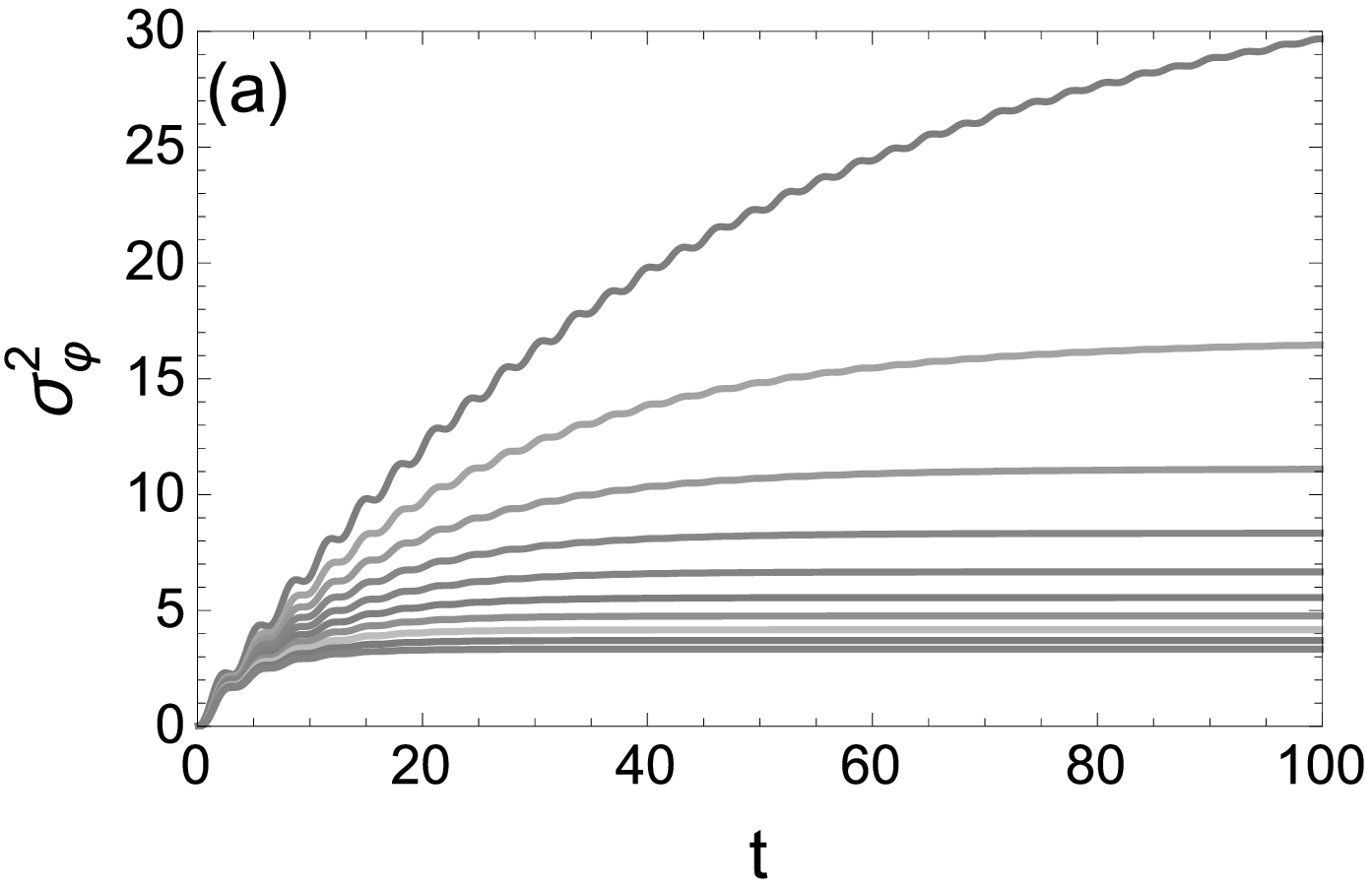}
    \includegraphics[width=0.4125\textwidth]{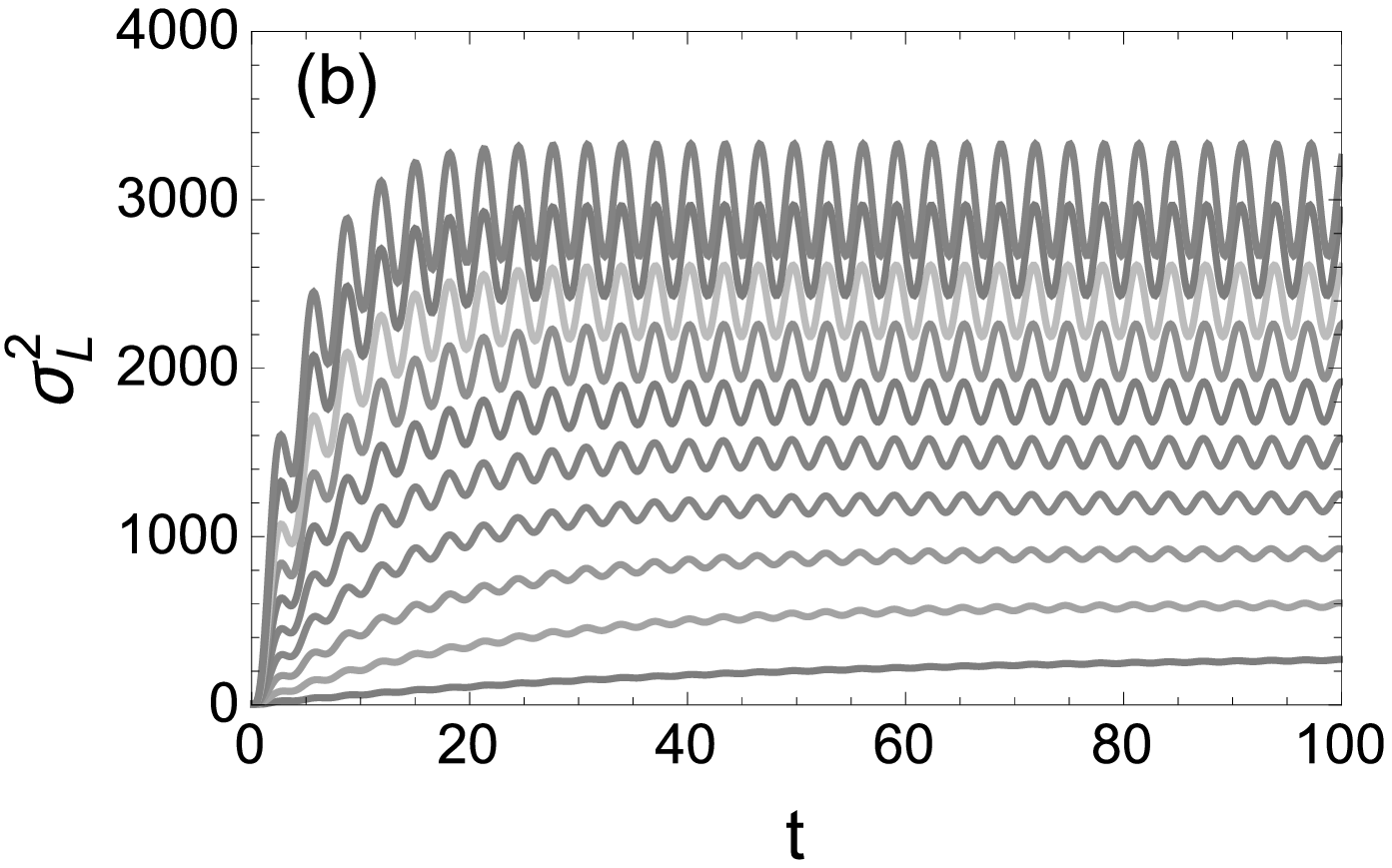}
\caption{The parameters values $\gamma_{\circ}=0.011$ and $k_BT=100$ are used
for the square of the standard deviation for: (a) the angle (the top line for $N=1$ and the bottom line for $N=10$, consecutively) and (b) the angular momentum (the top line for $N=10$ and the bottom line for $N=1$, consecutively).}
\end{figure*}

\noindent {\it (b)} For $k_BT=0.01$ for all values of $\gamma_{\circ}$, we observe {\it decrease of both} $\Delta\hat\varphi$ and $\Delta\hat L_z$ with time, Figure 9. Dependence on the number $N$ of blades is expected: $\Delta\hat\varphi$ decreases while $\Delta\hat L_z$ increases with the increase of $N$.
Interestingly, the decrease (damping) of the oscillation amplitude is different for the rotators of different sizes. In Figure 9,
the maximum damping is around $N=7,8$ for short times for both $\Delta\hat\varphi$ and $\Delta\hat L_z$,
while for $\Delta\hat L_z$ the maximum is around $N=3,4$ for longer time intervals. From Figure 10 we can detect the maximum damping for the angle as in Figure 9, but for the angular momentum standard deviation, the maximum damping is around $N=3$.

\begin{figure*}[!ht]
\centering
    \includegraphics[width=0.4\textwidth]{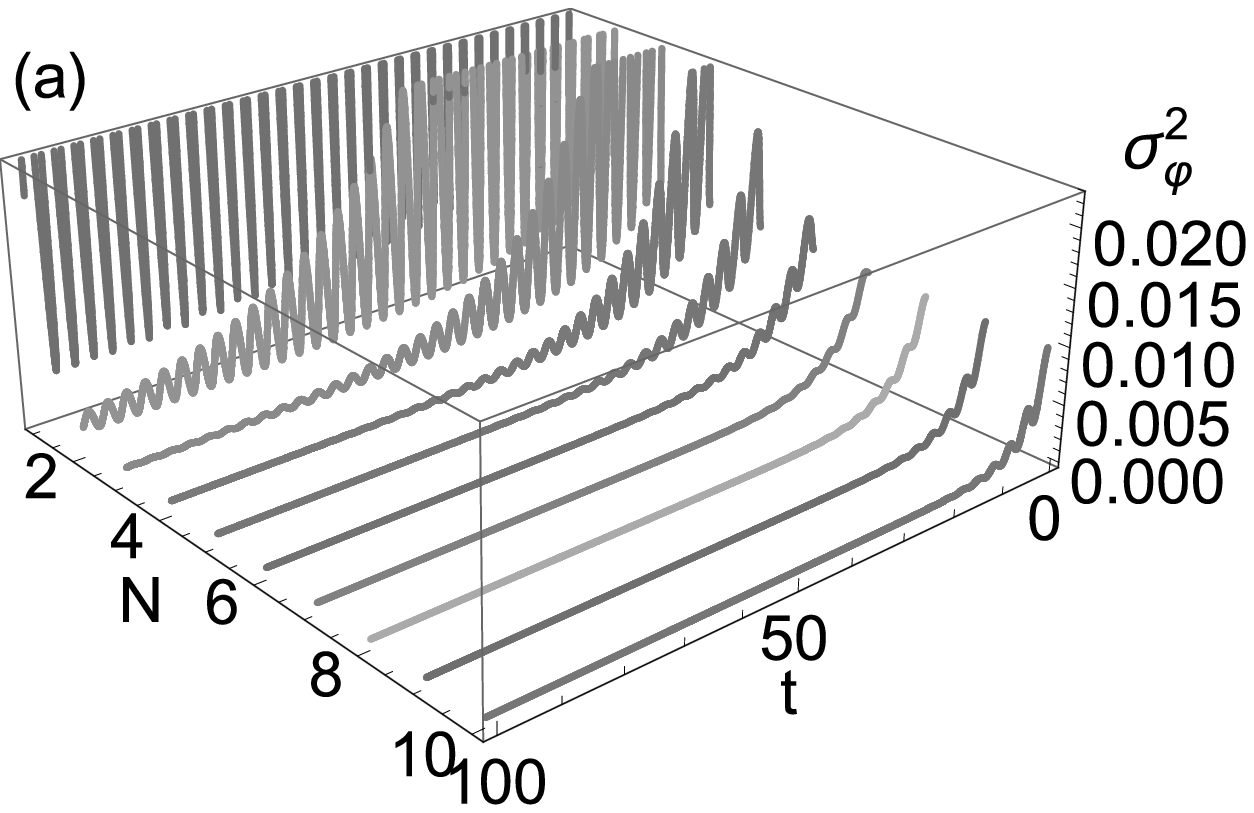}
    \includegraphics[width=0.4\textwidth]{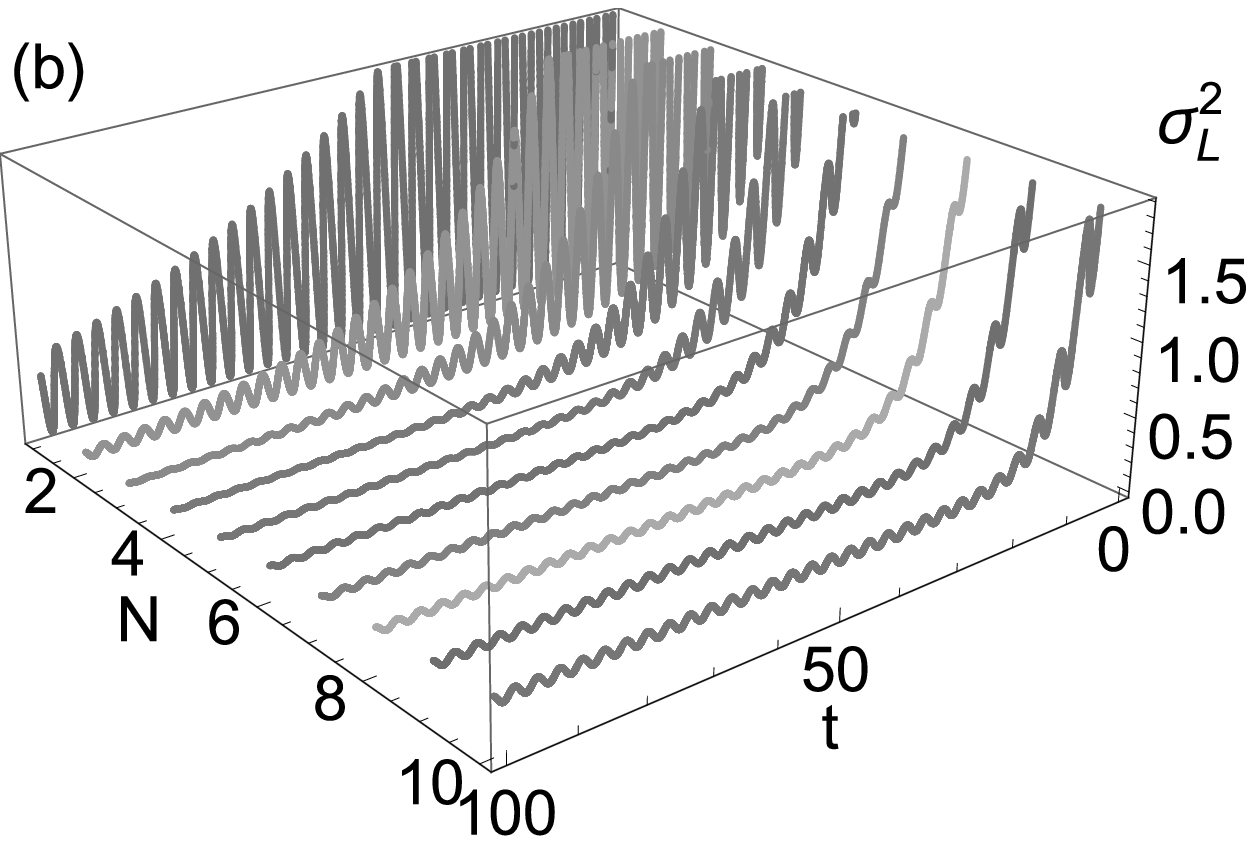}
\caption{The parameters values $\gamma_{\circ}=0.011$ and $k_BT=0.01$ are used
for the square of the standard deviation for: (a) the angle and (b) the angular momentum.}
\end{figure*}

\noindent {\it (c)} For the medium temperature, e.g. $k_BT=0.1$, it is observed saturation for the time-change of both $\Delta\hat\varphi$ and $\Delta\hat L_z$ thus exhibiting a smooth transition between the two cases {\it (a)} and {\it (b)}, while the minimum observed for the case {\it (b)} is now shifted to the values $N=2-4$--cf. Figure 10.

\begin{figure*}[!ht]
\centering
    \includegraphics[width=0.4\textwidth]{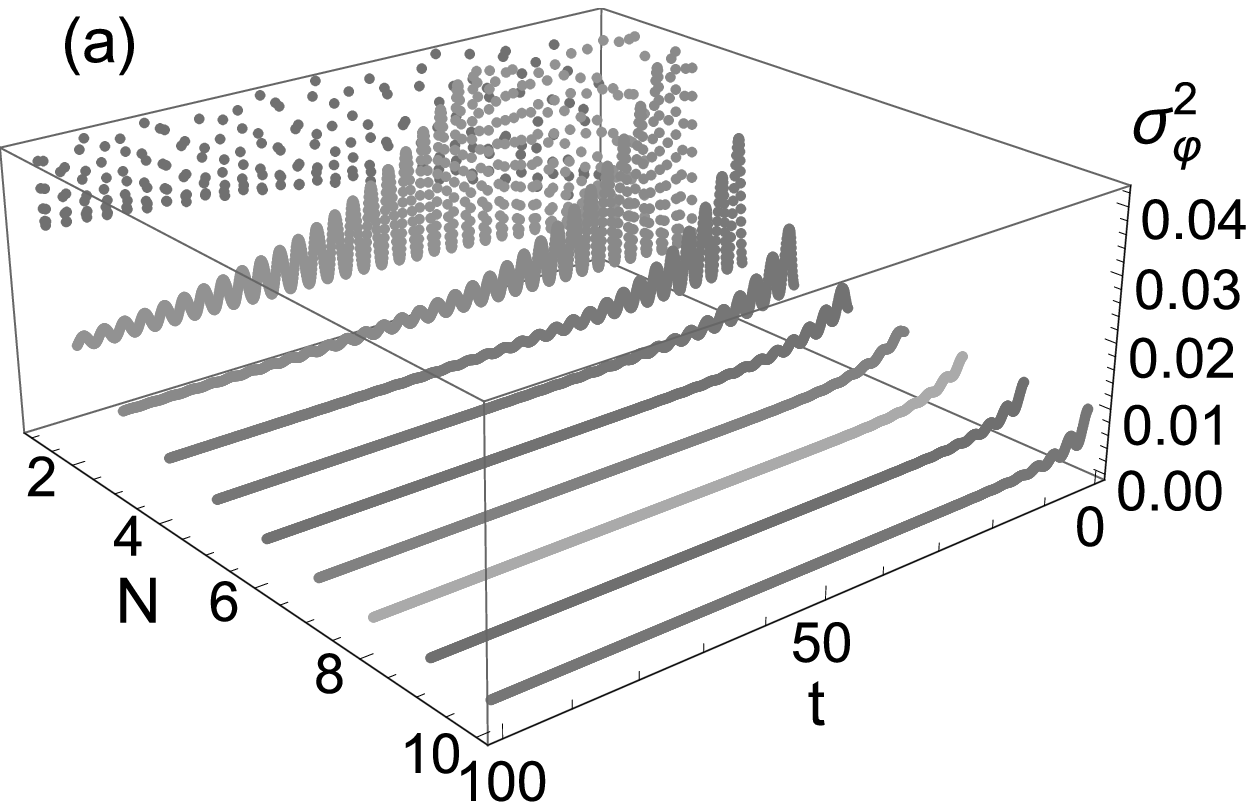}
    \includegraphics[width=0.4\textwidth]{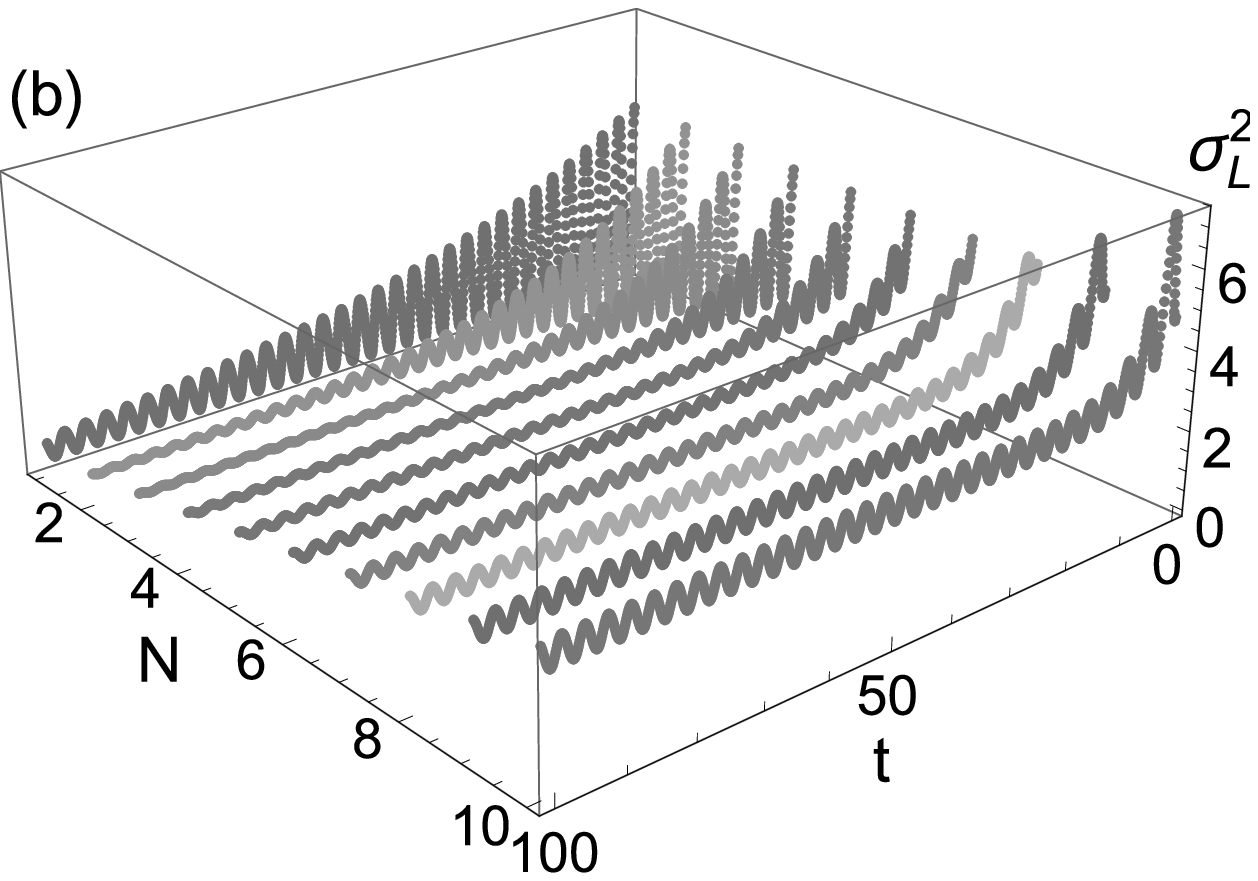}
\caption{The parameters values $\gamma_{\circ}=0.011$ and $k_BT=0.1$ are used
for the square of the standard deviation for: (a) the angle and (b) the angular momentum.}
\end{figure*}

\subsection{Comments}

The ''regular'' (expected) behavior found for the pure harmonic rotator \cite{JJD1} is here detected practically only for the high temperature cases that are presented by Figures 5 and 8: both $\Delta\hat\varphi$ and $\Delta\hat L_z$ increase, more-or-less, monotonically with time, while $\Delta\hat\varphi$ decreases and $\Delta\hat L_z$ increases with the increase of the number $N$ of the blades.
For all other combinations of the parameters, importance of the small nonharmonic term appears as follows: (i) Unexpectedly, a time {\it decrease} of the standard deviations for both the angle and  angular momentum is found as presented by Figures 6 and 9; (ii) In certain cases, e.g. Figure 6(b), a sharp initial decrease and nonmonotonic behavior of $\Delta\hat L_z$ is obtained; (iii) A non-monotonic dependence of $\Delta\hat\varphi$ on the number of blades $N$ is observed, cf. Figure 6(a); (iv) In certain cases (e.g., Figures 5, 8 and 9) the one-blade rotator exhibits large instability i.e. a large and fast increase of $\Delta\hat\varphi$ for short time; (v) Damping of the oscillation amplitude  exhibits a nonmonotonic dependence on the number of blades, with the minimum values depending on the temperature-see Figures 9 and 10.

It is important to stress that for the medium values of $k_BT$, a transition from the "regular" (the cases {\it (a)}, i.e. large $k_BT$) to the unexpected (the cases {\it (b)}, i.e. small $k_BT$) behavior is detected. That is, in Figures 7 and 10, there is a relatively fast (at least approximate) saturation of the standard deviations, without further decrease or increase for both $\Delta\hat\varphi$ and $\Delta\hat L_z$. Therefore we conclude that the above point (i) is not a "pathology" or a result of incorrect numerics. Rather, we observe a consistent dynamics for  both  observables.

The magnitudes of the standard deviation for $\Delta\hat L_z$ are by orders larger than for $\Delta\hat\varphi$, for both regimes. For example, from Section V.A we can learn that the magnitude for the angle observable takes the values (approximately) from the interval $(0.035,30)$, while for the angular momentum the interval is $(0.3,3000)$; and similarly for Section V.B.
Therefore, dynamics of the angular momentum may be regarded much more unstable than dynamics of the angle observable; nevertheless, this should be kept in conjunction with the observation (see above) of large instability of the angle observable in certain cases for $N=1$.

\section{Discussion}

The concept of {\it size} of a system is poorly defined and investigated in the standard quantum theory. It is not only a matter of  number of the constituent particles in the system but also of the specific choice of
degrees of freedom that describe the system's geometrical configuration or shape. This is still an open issue of the general quantum theory of open systems \cite{zurek, JJD2,JJD3}. In this paper we do not tackle the issue of the microscopic quantum origin of the definite size and shape of the composite quantum systems. Rather, we assume the propeller-like shape of certain molecular-rotators species as a {\it phenomenological data}, which is used in our considerations. Fortunately, introducing the size for the propeller-shaped rotators is possible \cite{JJD1} and is used as the starting point of the present study. Linear dependence of the damping factor $\gamma$ and the moment of inertia $I$ on the number $N$ of the blades makes these two parameters mutually dependent on each other--in contrast to the standard theory \cite{breuer,CL}.

We restrict our considerations to the maximum  $N=10$ blades for at least  two reasons. On the one hand, this is in accordance with the present state of the art in producing the  molecular rotators \cite{kottas}. On the other hand, increase in the number of the blades results in the decrease of the size of the environment monitoring the individual blades, thus possibly jeopardizing the assumption of the sufficiently large environment for every blade separately. Finally, for very large $N$, the propeller becomes similar to the rotating disc, which is a completely different model.

Limitations of our considerations follow from the choice of the method of Caldeira and Leggett as well as from the choice of investigation of the first passage time and the dynamics of the standard deviations for the conjugate observables of the rotator.  As it is emphasized in Introduction, results regarding the first passage time cannot be straightforwardly used for estimation of the escape rates even for the same physical model of the rotator. To this end a separate analysis of both the escape time \cite{kramers,melnikov} as well as of the first passage time on the basis of a semiclassical Wigner master equation \cite{coffey1,coffey2,coffey3,coffey4} can be recognized as another direction of the future research worth pursuing.

Our definition of the quantum first-passage-time does not coincide with those already used in the literature. In Ref. \cite{tao}, the time needed for the transition regarding the well-defined initial and the final state has been investigated with the mean-FPT and calculated for all the possible intermediate transitions. This method allows for the calculation of the FPT for the first moments of all the system observables but not including the continuous-variable (CV) systems, which is our case.
The quantum random walk model reduces to the classical one in  case of the one-dimensional CV system \cite{pawela}--which is our case. However, this does not allow for the direct comparison of our results with the classical counterparts, including the NES effect \cite{deck,dubkov,spagnolo1,spagnolo2,spagnolo3}. In this regard, on the one hand, quantum formalism does not allow for the well-defined spatial trajectories of the quantum system. On the other hand, the QFPT, introduced in this paper, even in the classical context, does not reveal much about the mean-FPT. That is, when the average (mean) value of the relevant variable attains the threshold value, there are  the classical trajectories well above as well as below the threshold value. Therefore, in general, the concept of the mean-FPT is well defined only in the classical-physics context. Its transfer to the quantum-mechanical context, especially while introducing the concept of {\it size} of the quantum objects is non-trivial and here not fully elaborated.

The two methods used in Sections IV and V qualitatively coincide but are not mutually equivalent. As emphasized in Introduction, the results presented in Sections IV and V do not qualitatively change with the variations of the parameters $\omega$ and $b$; those variations indirectly include the possible dependence of both $\omega$ and $b$ on the number $N$ of the propeller blades.

The QFPT method is suitable for investigating the system dynamics on the very short time scale that does not exhibit any significant role of certain system parameters, such as  environment temperature $k_BT$ or the initial position $\varphi_{\circ}$. Stability of the rotation increases with the increase of the number $N$ of the blades that, in turn, proves {\it not} to be reducible to the system inertia.

In addition, the standard deviations of both the angle and  angular momentum observables exhibit strong parameter-dependence with some unexpected behavior as presented in Sections IV.C and V.C. None of these findings appear for the purely harmonic case \cite{JJD1}, for which
monotonic increase with time applies for both $\Delta\hat\varphi$ and $\Delta\hat L_z$, with the general decrease of $\Delta\hat\varphi$ and the increase of $\Delta\hat L_z$ with the increase of the number $N$ of the blades; this  behavior can be found in Figures 5 and 8 as well as in Figure 7 for the $\hat L_z$ observable for longer time intervals.

Our results include the standard overdamped ($\gamma_{\circ} > k_BT$) and underdamped ($10\gamma_{\circ} < k_BT$) regimes that are presented by Figures 6 and 7, and by the Figures 5 and 8, respectively.  Compared in this context, Figures 6 and 7 for the overdamped regime reveal the unexpected decrease of the standard deviations for both observables.  The decrease is more pronounced for the angle than for the angular-momentum observable, while, on the other hand, it is more pronounced for the lower temperature. The behavior obtained for the underdamped regime, presented by Figures 5 and 8 exhibits the initial increase of the standard deviations for both observables  that is followed by saturation in the longer time intervals, for every number of blades.

Now, borrowing from Section V.C, the list of the stability criteria for the standard deviations known for the purely harmonic case \cite{JJD1} is nontrivially extended and varied as follows.

(A) The choice of the observable to be acted on.

\noindent {\it Comments}. The choice of the observable to be externally manipulated strongly depends on the criteria (B)-(E) presented below, while bearing in mind that for certain cases (see above) dependence on the number $N$ of the blades provides additional contributions that differ for the conjugate observables $\hat\varphi$ and $\hat L_z$.  E.g., manipulating the angle may be preferable for the case $\gamma_{\circ}/\omega>1$ for the environment on the low temperature and for the relatively longer time intervals (after the initiation/preparation of the rotator), when the choice of $N\in\{4,5,6\}$ should be made especially if the relatively small magnitude of change of the standard deviation is required--cf. Figure 6(a). However,
if it is preferable to quickly perform the fast actions that are to be exerted on the system (cf. the criterion (E))
after the system initialization, then manipulation of the angular momentum may be a preferred choice when $N\in\{3,4,5\}$ should be made - cf. Figure 6(b).

(B) The parameter regime.

\noindent {\it Comments}. Even for the same ratio of $10\gamma_{\circ}/\omega$,  different behavior of the standard deviations is observed--compare Figure 5 with Figure 6 (i.e. compare Figure 8 with Figure 9). Generally, the small temperature of the environment provides better stability however, in certain cases there are exceptions referring to
the short time behavior--which (cf. the criteria (D) and (E) below) is of importance for the protocols right after the system initialization.

(C) The magnitude of change of the standard deviations.

\noindent {\it Comments}. Typically, the magnitude for the angle is smaller, except for $N=1$ (cf. e.g. Figure 5). The different conclusions are drawn when the amplitude of oscillation is in question. Then a nonmonotonic dependence on the number $N$ of the blades is found--cf. Figures 9 and 10.

(D) The short versus the long time behavior.

\noindent {\it Comments}. The short time behavior is in strong conjunction with the above items (A) and (B) and may prefer the relatively high temperature--compare e.g. Figures 5 and 8.
The long-time behavior generally exhibits saturation of the standard deviations and, in this sense, a more reliable prediction. Needless to say, the choice of the time scale for the system manipulation cannot be made without a reference to all the other criteria, notably the criterion  (E). Additionally, for the longer time intervals, the nonmonotonic dependence of $\Delta\hat\varphi$ on $N$ is observed--see Figure 6.

(E) The rate of the external actions.

\noindent {\it Comments}. It can be expected \cite{JJD1} that the external actions that are not included in the master equation (2) can increase (and possibly accumulate)  standard deviations for both choices in (A). Therefore a large number of  fast actions performed in a short time interval may lead to the uncontrollable increase in the standard deviation(s) as compared with the small number of the longer lasting actions in the same time interval. Thus usefulness of the quick versus the slow actions is in strong conjunction with the above criteria, notably with the criterion (D).

Therefore we face the absence of  simple rules or recipes for designing the protocols for
 desired control of the propeller-shaped molecular (rigid) planar rotators. The possible
combinations of the criteria require a procedure that is along the lines of optimization procedures
in engineering \cite{ravindran}. This requires a separate and careful analysis that is not part of this paper.
This constitutes our answer to the question posed in Introduction on the physical role of
the size and shape for the dynamical stability of the propeller-shaped molecular rotators.
As a noteworthy part of the answer, we stress the fact that the role of
the size of the propellers cannot be reduced to the more-or-less pure inertial effects widely
known and expected in the classical physical context.

\section{Conclusion}

The presence of a small cubic term in the external potential for the rotator introduces significant departure from the exact harmonic potential. Particularly, the standard deviations for the angle and  angular momentum observables may dynamically decrease for some parameter regimes while exhibiting nonlinear dependence on the number of "blades" of the propeller-like shaped molecular rotator. We also observe irreducibility of the obtained results to the purely inertial effects, which may be intuitively expected for the classical regime of the rotator dynamics. The sensitivity of  rotation to  details of the model and the parameter regimes emphasizes that utilizing the propeller rotations stability is an optimization problem that requires a separate careful analysis.

\appendix*\section{Differential equations for the moments}

Derivation of the differential equations for the moments, eq.(3), is straightforward but rather tedious. Here we provide the exact results regarding eq.(4), up to the moments of the fourth order.

The transposed vector composed of the moments:
$X^T = \{ \langle \hat\varphi \rangle, \langle \hat L_z \rangle, \langle \hat \varphi^2 \rangle,$    $\langle \hat\varphi \hat L_z + \hat L_z \hat\varphi \rangle, \langle \hat L_z^2 \rangle,
\langle \hat\varphi^3\rangle, \langle \hat\varphi^2\hat L_z + \hat L_z\hat\varphi\rangle, \langle \hat\varphi \hat L_z^2 + \hat L_z^2 \hat\varphi \rangle, \langle \hat L_z^3 \rangle, \langle \hat\varphi^4 \rangle,
\langle \hat\varphi^3\hat L_z + \hat L_z\hat\varphi^3\rangle$, $\langle \hat\varphi^2 \hat L_z^2 + \hat L_z^2\hat\varphi^2\rangle, \langle  \hat\varphi \hat L_z^3 + \hat L_z^3\hat\varphi \rangle,  \langle \hat L_z^4\rangle
\}$.

The related matrix $\mathcal{M}$ reads:

\begin{equation}
\left({
  \begin{array}{cccccccccccccc}
    0       & \alpha_1 & 0 & 0 & 0 &  0 & 0 & 0 & 0 & 0 & 0 & 0 & 0 & 0  \\
    -\alpha_2       & -2\alpha_3 & -3b &  0 &  0 & 0 & 0 & 0 & 0 & 0 & 0 & 0 & 0 & 0  \\
    0 &0 &  0 & \alpha_1 &  0 & 0 & 0 &  0 & 0 & 0 & 0 & 0 & 0 & 0  \\
    0 & 0 & -2\alpha_2 & -2\alpha_3 &  2\alpha_1 & -6b  & 0 & 0 & 0 &  0 & 0 & 0 & 0 & 0    \\
    0 & 0 & 0 & -\alpha_2 & -4\alpha_3 & 0 & -3b & 0 & 0 &  0 & 0 & 0 & 0 & 0  \\
        0 & 0 & 0 & 0 & 0 & 0 & {3\alpha_1\over 2} & 0 & 0 &  0 & 0 & 0 & 0 & 0  \\
        0 & 0 & 0 & 0 &  0 & -2\alpha_2 & -2\alpha_3 & 2\alpha_1 & 0 &  -6b & 0 & 0 & 0 & 0 \\
        8C_1 & 0 & 0 & 0 & 0 & 0 & -2\alpha_2 & -4\alpha_3 & 2\alpha_1 &  0 & -6b & 0 & 0 & 0\\
        0 & 12 C_1 & 0 & 0 & 0 & 0 & 0 & {-3\alpha_2\over 2} & -6\alpha_3 &  0 & 0 & {-9b\over 2} & 0 & 0\\
        0 & 0 & 0 & 0 & 0 & 0 & 0 & 0 & 0 &  0 & 2\alpha_1 & 0 & 0 & 0\\
        0 & 0 & 0 & 0 & 0 & 0 & 0 & 0 & 0 &  -2\alpha_2 & -2\alpha_3 & 3\alpha_1 & 0 & 0\\
        0 & 0 & 8C_1 & 0 & 0 & 0 & 0 & 0 & 0 &  0 & -2\alpha_2 & -4\alpha_3 & 2\alpha_1 & 0\\
        -24b\hbar^2 & 0 & 0 & 12C_1 & 0 & 0 & 0 & 0 & 0 &  0 & 0 & -3\alpha_2 & -6\alpha_3 & 2\alpha_1\\
       0 &  -12b\hbar^2 & 0 & 0 & 24C_1 & 0 & 0 & 0 &  0 & 0 & 0 & 0 & -2\alpha_2 & -8\alpha_3\\
  \end{array}}
\right)
\end{equation}

\noindent while the transposed vector $K$:

\begin{eqnarray}
&\nonumber&
K^T = \{0,0,0,0, 4C_1, 0,0,0,-3b\hbar^2, 0, 3\alpha_1 \hbar^2-3b \langle \hat\varphi^5\rangle, -4\alpha_3\hbar^2\\&&
- 6b \langle \hat\varphi^4 \hat L_z + \hat L_z \hat\varphi^4 \rangle, -3\alpha_2 \hbar^2 -
9b \langle \hat\varphi^3 \hat L_z^2 + \hat L_z^2 \hat\varphi^3 \rangle,-6b \langle \hat\varphi^2 \hat L_z^3 + \hat L_z^3 \hat\varphi^2\rangle\}.
\end{eqnarray}

\noindent In eqs.(A.1) and (A.2): $\alpha_1 = 1/I, \alpha_2 = I\omega^2, \alpha_3 = \gamma, C_1 = I\gamma k_BT$.

From eq.(A.1) it is obvious that, for the cubic potential, the set of  equations is {\it not closed}: a set of equations for one order of the moments depends on the higher-order moments (cf. the terms proportional to the constant $b$). For the exactly harmonic potential ($b=0$), the sets of the equations are closed for every order of the moments.
The expressions for the first moments for the purely harmonic case are well known to read:

\begin{eqnarray}
&\nonumber&
\langle\hat\varphi\rangle_{b=0}  = \exp(-\gamma t) \left(
\langle \hat\varphi(0) \rangle \left(\cosh \Omega t + {\gamma\over\Omega} \sinh \Omega t  \right) + {\langle \hat L_z(0)\rangle \over I\Omega} \sinh \Omega t
\right)
\\&&
\langle\hat L_z\rangle_{b=0}   = \exp(-\gamma t) \left(
\langle \hat L_z(0) \rangle \left(\cosh \Omega t - {\gamma\over\Omega} \sinh \Omega t  \right) - {I\omega^2\over \Omega} \langle \hat \varphi(0)\rangle \sinh \Omega t
\right)
\end{eqnarray}

\noindent while for the standard deviations we borrow the exact solutions from \cite{JJD1}:

\begin{eqnarray}
&\nonumber&
(\Delta\hat\varphi)^2_{b=0} =
{k_BT\over I\omega^2 \Omega^2} \left(\Omega^2 + \exp(-2\gamma t) (\omega^2 - \gamma^2 \cosh(2\Omega t) -\gamma\Omega\sinh(2\Omega t)) \right)
+\\&&\nonumber
{\langle \Delta \hat L_z(0)\rangle^2\over I^2\Omega^2} \exp(-2\gamma t) \sinh^2(\Omega t)
+\\&&\nonumber
{\langle \Delta \hat\varphi(0)\rangle^2\over \Omega^2} \exp(-2\gamma t) \left(
-\omega^2 \cosh^2(\Omega t) + \gamma^2 \cosh(2\Omega t) + \gamma\Omega \sinh(2\Omega t)
\right)
+\\&&\nonumber
{\sigma_{\varphi L}(0) \over 2I\Omega^2} \exp(-2\gamma t) \left(
2\gamma \sinh^2(\Omega t)  + \Omega \sinh(2\Omega t)
\right),
\\&&\nonumber
(\Delta \hat L_z)^2_{b=0}  = {Ik_BT\over\Omega^2} ( -\omega^2 (1- \exp(-2\gamma t)) +\gamma^2 (1-\exp(-2\gamma t)\cosh(2\Omega t))\\&&\nonumber
- \gamma\Omega\exp(-2\gamma t) \sinh(2\Omega t))\\&&\nonumber
+
{(\Delta\hat L_z(0))^2\over \Omega^2} \exp(-2\gamma t) \left(
-\omega^2\cosh^2(\Omega t) + \gamma^2 \cosh(2\Omega t) - \gamma\Omega\sinh(2\Omega t)\right)\\&&\nonumber
+
{I^2\omega^4\over\Omega^2} (\Delta\hat\varphi(0))^2 \exp(-2\gamma t) \sinh^2(\Omega t)\\&&
+
{I\omega^2\over 2\Omega^2} \sigma_{\varphi L}(0) \exp(-2\gamma t) \left(
2\gamma \sinh^2(\Omega t) - \Omega\sinh(2\Omega t)
\right).
\end{eqnarray}

\noindent where $\Omega = \sqrt{\gamma^2 - \omega^2}$ and $\sigma_{\varphi L} = \langle \hat \varphi\hat L_z + \hat L_z\hat\varphi\rangle-2\langle\hat\varphi\rangle\langle\hat L_z\rangle$.
The expressions for $\langle\hat\varphi^2\rangle$ and $\langle\hat L_z^2\rangle$ are obtained from equations (B.1)-(B.3) in \cite{JJD1} by replacing $(\Delta\hat\varphi)^2$ by $\langle\hat \varphi^2\rangle$, the $(\Delta\hat L_z)^2$ by $\langle\hat L_z^2\rangle$, and $\sigma_{\varphi L}$ by $\langle\hat\varphi \hat L_z + \hat L_z\hat\varphi\rangle$.

From equation (A.1) follows the set of the  equations for the third-order moments for the exact harmonic potential ($b=0$):

\begin{eqnarray}
&\nonumber&
{d\langle\hat\varphi^3\rangle_{b=0} \over dt}  = {3\over 2I}\langle\hat\varphi^2 \hat L_z +\hat L_z\hat\varphi^2\rangle_{b=0},
\\&&\nonumber
{d\langle \hat\varphi^2\hat L_z + \hat L_z\hat\varphi^2\rangle_{b=0} \over dt} = -2I\omega^2 \langle\hat\varphi^3\rangle_{b=0} - 2\gamma \langle \hat\varphi^2\hat L_z + \hat L_z\hat\varphi^2\rangle_{b=0}+ {2\over I} \langle \hat\varphi\hat L_z^2 + \hat L_z^2\hat\varphi\rangle_{b=0},
\\&&\nonumber
{d\langle \hat\varphi \hat L_z^2 + \hat L_z^2\hat\varphi\rangle_{b=0} \over dt} = -2I\omega^2 \langle \hat\varphi^2\hat L_z + \hat L_z\hat\varphi^2\rangle_{b=0} - 4\gamma \langle \hat\varphi \hat L_z^2 + \hat L_z^2\hat\varphi\rangle_{b=0} + {2\over I}\langle \hat L_z^3\rangle_{b=0}\\&&\nonumber + 8I\gamma k_BT\langle\hat\varphi\rangle_{b=0},
\\&&
{d\langle\hat L_z^3\rangle_{b=0} \over dt}  = - {3\over 2}I\omega^2 \langle \hat\varphi \hat L_z^2 + \hat L_z^2\hat\varphi\rangle_{b=0} - 6\gamma \langle \hat L_z^3\rangle_{b=0} + 12 I\gamma k_BT \langle\hat L_z\rangle_{b=0}.
\end{eqnarray}

With the use of eq.(A.3), the system eq.(A.5) becomes closed.
Analytical solutions of eq.(A.5) are rather large and physically nontransparent, and therefore  will not be explicitly given here. Solutions for  $\langle \hat\varphi^3\rangle_{b=0}$
and $\langle \hat\varphi \hat L_z + \hat L_z \hat\varphi^2 \rangle_{b=0}$ are implicit to our numerical calculations performed in  Section IV.

%\section*{Acknowledgement}

\begin{acknowledgments}
{This research has been performed as a part of the work within the project No.~171028 funded by the Ministry of Education, Science and Technological Development of the Republic of Serbia. We thank Nevena Bankovi\' c for discussions and the anonimous referees for useful remarks that improved presentation of our results.}
\end{acknowledgments}

%\section*{References}

\end{document}